\documentclass[12pt, a4paper]{article}
\usepackage{graphicx}
\usepackage{amssymb}
\usepackage{amsmath}
\usepackage{bm}
\usepackage{color}
\usepackage{theorem}
\usepackage{subcaption}
\usepackage{listings}
\usepackage{colortbl}
\usepackage{tabularx}
\usepackage{longtable}
\usepackage[utf8]{inputenc}
\usepackage[T1]{fontenc}
\usepackage{lmodern}
\usepackage[top=30truemm,bottom=30truemm,left=25truemm,right=25truemm]{geometry}

\usepackage[sort&compress,numbers, merge]{natbib}

\definecolor{Orange}{cmyk}{0,0.61,0.87,0}
\definecolor{JungleGreen}{cmyk}{0.99,0,0.52,0}
\definecolor{OliveGreen}{cmyk}{0.64,0,0.95,0.40}
\definecolor{Brown}{cmyk}{0,0.81,1,0.60}
\definecolor{RoyalBlue}{cmyk}{0.71,0.53,0,0.12}
\definecolor{Gray}{cmyk}{0,0,0,0.40}
\definecolor{LightPink}{cmyk}{0.0,0.25,0,0}
\definecolor{LLightPink}{cmyk}{0.0,0.10,0,0}
\definecolor{LightBlue}{cmyk}{0.25,0,0,0}
\definecolor{LightGray}{cmyk}{0,0,0,0.2}

\newcommand{\1}{\mbox{1}\hspace{-0.25em}\mbox{l}}

\allowdisplaybreaks[1]

\usepackage[colorlinks=true, linkcolor=OliveGreen, citecolor=RoyalBlue,
urlcolor=RoyalBlue]{hyperref}

\renewcommand{\thefootnote}{\fnsymbol{footnote}}

\begin{document}

\begin{titlepage}

\begin{flushright}
  %   {\tt
  %     UT-18-**
  %   }

\end{flushright}

\vskip 1.35cm
\begin{center}

%\textcolor{RoyalBlue}
{\Large
{\bf
Minimal Nambu-Goldstone Higgs Model \\[3pt]
in Supersymmetric SU(5) Revisited 
}
}

\vskip 1.5cm

Koichi Hamaguchi$^{a,b}$\footnote{
  E-mail address: \href{mailto:hama@hep-th.phys.s.u-tokyo.ac.jp}{\tt hama@hep-th.phys.s.u-tokyo.ac.jp}}, 
Shihwen Hor$^a$\footnote{
  E-mail address: \href{mailto:shihwen@hep-th.phys.s.u-tokyo.ac.jp}{\tt shihwen@hep-th.phys.s.u-tokyo.ac.jp}},
Natsumi Nagata$^a$\footnote{
E-mail address: \href{mailto:natsumi@hep-th.phys.s.u-tokyo.ac.jp}{\tt natsumi@hep-th.phys.s.u-tokyo.ac.jp}}

\vskip 0.8cm

{\it $^a$Department of Physics, University of Tokyo, Bunkyo-ku, Tokyo
 113--0033, Japan} \\[2pt]
 {\it ${}^b$Kavli Institute for the Physics and Mathematics of the
 Universe (Kavli IPMU), University of Tokyo, Kashiwa 277--8583, Japan}

\date{\today}

\vskip 1.5cm

\begin{abstract}

We revisit the minimal Nambu-Goldstone (NG) Higgs supersymmetric (SUSY) SU(5) grand unified model and study its phenomenological implications.
The Higgs sector of the model possesses a global SU(6) symmetry, which is spontaneously broken and results in the Higgs doublets of the minimal SUSY Standard Model (MSSM) as NG chiral superfields. Therefore, the model naturally leads to light Higgs doublets and solves the doublet-triplet splitting problem.
Because of the SU(6) symmetry, the couplings of the Higgs sector are tightly restricted, and thus the model is more predictive than the minimal SUSY SU(5).
We determine all the grand-unified-theory parameters via the matching conditions of the gauge coupling constants at the unification scale and calculate proton lifetime, confronting this with current experimental bounds. We discuss that this model is incompatible with the constrained MSSM, whilst it has a large viable parameter space in the high-scale SUSY scenario. The perturbativity condition on the trilinear coupling of the adjoint Higgs field imposes an upper (lower) limit on the wino (gluino) mass, implying a hierarchical mass pattern for these gauginos. Future proton-decay searches can probe a large part of the parameter space, especially if the SUSY-breaking scale is $\lesssim 100$~TeV. 

\end{abstract}

\end{center}
\end{titlepage}

\renewcommand{\thefootnote}{\arabic{footnote}}
\setcounter{footnote}{0}

%%%%%%%%%%%%%%%%%%%%%%%%%%%%%%%%%%%%%%%%%%
\section{Introduction}
\label{sec:introduction}
%%%%%%%%%%%%%%%%%%%%%%%%%%%%%%%%%%%%%%%%%%

%%%%Introduction about what GUT is%%%%%%%%%%%%
Unification of interactions has been one of the main goals of particle physics. The minimal supersymmetric (SUSY) SU(5) grand unified theory (GUT)~\cite{Dimopoulos:1981zb,Sakai:1981gr} offers a desirable framework to this end, where the three Standard Model (SM) gauge couplings are unified with a great accuracy~\cite{Dimopoulos:1981yj, Marciano:1981un, Einhorn:1981sx, Amaldi:1991cn, Langacker:1991an, Ellis:1990wk, Giunti:1991ta}. Nevertheless, the minimal SU(5) GUT is known to have a fatal defect, called the doublet-triplet splitting problem. In SU(5), the Higgs doublets in the minimal SUSY SM (MSSM) are embedded into $\mathbf{5}$ and $\overline{\mathbf{5}}$ representations, accompanied with color-triplet components. These color-triplet fields induce proton decay, and to evade the limits imposed by proton-decay searches they must have a GUT-scale mass. On the other hand, the MSSM Higgs doublets need to have a SUSY-scale mass to achieve a successful electroweak-symmetry breaking. This mass splitting is realized with a huge amount of fine-tuning in the minimal SU(5), making this model less appealing. 

%%% NG Higgs GUT %%%
An attractive idea to solve this doublet-triplet splitting problem is that the Higgs boson is a pseudo-Nambu-Goldstone (pNG) boson associated with the spontaneous breaking of a global symmetry. Such a scenario in the framework of GUTs was first explored in Ref.~\cite{Inoue:1985cw} and discussed later in Refs.~\cite{Anselm:1986um, Anselm:1988ss, Berezhiani:1989bd, Inoue:1990ac, Barbieri:1992yy, Barbieri:1993wz, Barbieri:1994kw, Berezhiani:1995sb, Berezhiani:1995dt, Csaki:1995cf, Dvali:1996sr, Shafi:1998jf, Shafi:1999tn, Bajc:2001zz, PaccettiCorreia:2002lpe, Bando:2002vy, CarcamoHernandez:2010xgi, Tavartkiladze:2018rrq}. In the model considered in Ref.~\cite{Inoue:1985cw}, the $\mathbf{5}$ and $\overline{\mathbf{5}}$ Higgs representations, as well as an adjoint Higgs field of SU(5), are embedded into an adjoint representation of an SU(6) global symmetry. The SU(5) GUT gauge group is a subgroup of this SU(6). The vacuum expectation value (VEV) of this adjoint field breaks both the SU(6) global symmetry and the SU(5) GUT gauge symmetry, giving masses to the SU(5) gauge bosons and yielding a pair of massless doublet Higgs fields as NG multiplets. The color-triplet components of the $\mathbf{5}$ and $\overline{\mathbf{5}}$ representations remain massive, \textit{i.e.}, have a GUT-scale mass. The mass term of the doublet Higgs fields is protected from quantum corrections thanks to the non-renormalization theorem in SUSY theories, and is generated through the SUSY-breaking effect. As a result, the MSSM Higgs doublets acquire a mass around the SUSY-breaking scale, and hence, the doublet-triplet splitting problem is solved in a natural manner.  We refer to this setup as the minimal NG Higgs SUSY SU(5) GUT model. 

%%%%What we did and the results%%%%%%%%%%%%%%%
In this work, we revisit this model and study its phenomenological implications in detail. Because of the global SU(6) symmetry in the Higgs sector, the number of free parameters in this model is smaller than that in the minimal SU(5), allowing us to determine all of the GUT parameters, such as the SU(5) gauge coupling constant, the trilinear coupling of the adjoint Higgs field, $\lambda$, the colored Higgs mass, $M_{H_C}$, and the GUT gauge boson mass, $M_X$, through the matching conditions of the gauge coupling constants at the GUT scale, $\alpha_a(Q_G)$ ($a=1,2,3$), where $Q_G$ is the unification scale defined by $\alpha_1 (Q_G) = \alpha_2 (Q_G)$. It is found from the perturbativity condition on $\lambda$ that $\alpha_2(Q_G) \gtrsim \alpha_3 (Q_G)$ and $M_{H_C} \lesssim M_X$ are favored. The former inequality restricts the low-energy SUSY mass spectrum since $\alpha_a(Q_G)$ depend on the masses of the MSSM SUSY particles through the renormalization group equations (RGEs). In addition, we can predict proton decay rates by determining $M_{H_C}$ and $M_X$. To show the significance of these results, we consider two scenarios for the SUSY mass spectrum: the constrained MSSM (CMSSM) and high-scale SUSY. It is found that the CMSSM is incompatible with the $p \to K^+\bar{\nu}$ bound from the Super-Kamiokande experiment~\cite{Super-Kamiokande:2014otb, Takhistov:2016eqm}, as the SUSY particles are predicted to lie around ${\cal O}(10)$~TeV in this case. In the high-scale SUSY scenario, on the other hand, we find a large viable parameter space. We also find that the perturbativity condition on $\lambda$ leads to an upper (lower) limit on the wino (gluino) mass, implying a hierarchical mass pattern for these gauginos. Future proton decay searches can test a large part of the viable parameter regions, especially if the SUSY-breaking scale is $\lesssim 100$~TeV. 

%%%%The organization of this paper%%%%%%%%%%%%
This paper is organized as follows. In Sec.~\ref{sec:model}, we review the minimal NG Higgs SUSY SU(5) GUT model and discuss its symmetry-breaking structure and mass spectrum. In Sec.~\ref{sec:matchingconditions}, we show how to extract the GUT parameters from the GUT-scale matching conditions of the SM gauge coupling constants. We also evaluate the mass parameters for the MSSM Higgs fields induced by the SUSY-breaking effect. Then, we show the results of our analysis for the CMSSM and high-scale SUSY in Sec.~\ref{sec:results}. Section~\ref{sec:summary} is devoted to conclusion and discussion. We summarize relevant formulae for the RGE analysis and proton decay calculation in Appendix~\ref{sec:RG} and \ref{sec:protondecay}, respectively.

%%%%%%%%%%%%%%%%%%%%%%%%%%%%%%%%%%%%%%%%%%
\section{Model}
\label{sec:model}
%%%%%%%%%%%%%%%%%%%%%%%%%%%%%%%%%%%%%%%%%%

The minimal NG Higgs SUSY SU(5) GUT model was first proposed in Ref.~\cite{Inoue:1985cw} and discussed later in Refs.~\cite{Anselm:1986um, Anselm:1988ss, Inoue:1990ac, Barbieri:1992yy, Csaki:1995cf}. In this setup, the Higgs sector is assumed to possess a global SU(6) symmetry, and the MSSM Higgs multiplets reside in the adjoint representation of the SU(6), $\hat{\Sigma}$. The SU(5) GUT gauge group is a subgroup of the SU(6), and hence, this global symmetry is explicitly broken by the gauge interaction. The global SU(6) symmetry is violated also by the couplings of the Higgs multiplets to the MSSM matter fields. The superpotential of this model thus has the structure 
\begin{equation}
  W = W_{\mathrm{Higgs}} (\hat{\Sigma}) + W_{\mathrm{Yukawa}} ~,
\end{equation}
where $W_{\mathrm{Higgs}} (\hat{\Sigma})$ respects the global SU(6) symmetry while $W_{\mathrm{Yukawa}}$, which includes the MSSM matter chiral superfields, does not. The SU(6)-symmetric part is 
\begin{equation}
  W_{\mathrm{Higgs}} (\hat{\Sigma}) = \frac{1}{3} \lambda \mathrm{Tr} \hat{\Sigma}^3 + \frac{1}{2} M \mathrm{Tr} \hat{\Sigma}^2 ~. 
\end{equation}

We decompose $\hat{\Sigma}$ in terms of SU(5) representations as 
\begin{equation}
  \hat{\Sigma} = 
  \begin{pmatrix}
    - 5 S /\sqrt{60}  & \bar{H}/\sqrt{2} \\ 
    {H}/\sqrt{2} &  S \1_5 /\sqrt{60} + \Sigma
  \end{pmatrix}
  ~, 
\end{equation}
where $\1_5$ is the $5\times 5$ identity matrix and $\Sigma \equiv \Sigma^A T^A$ with $T^A$ the SU(5) generators.\footnote{We normalize these generators as $\mathrm{Tr}(T^A T^B) = \delta^{AB}/2$. }  $W_{\mathrm{Higgs}} (\hat{\Sigma})$ is then expressed with these component fields as 
\begin{align}
  W_{\mathrm{Higgs}} (\hat{\Sigma}) &= \frac{1}{3} \lambda \mathrm{Tr} (\Sigma^3) + \frac{1}{2} \lambda \bar{H}\Sigma H + \frac{1}{2} M \mathrm{Tr} (\Sigma^2) + \frac{1}{2} M \bar{H}H \nonumber \\ 
  &- \frac{1}{3\sqrt{15}} \lambda S^3 -\frac{1}{\sqrt{15}} \lambda S \bar{H}H + \frac{1}{\sqrt{60}} \lambda S \mathrm{Tr} (\Sigma^2) 
  + \frac{1}{4} M S^2~.
  \label{eq:whiggs}
\end{align}
The terms in the first line appear also in the minimal SUSY SU(5)~\cite{Dimopoulos:1981zb, Sakai:1981gr}, where the coefficients of these terms are independent. On the contrary, there are relations among the coefficients in the present scenario, which play an important role in the following discussion. 

The adjoint Higgs $\hat{\Sigma}$ is assumed to have a VEV of the form 
\begin{equation}
  \langle \hat{\Sigma} \rangle = \hat{V} \cdot \mathrm{diag} (1,1,1,1,-2,-2) ~,
  \label{eq:vev}
\end{equation}
where $\hat{V} = M/\lambda$.\footnote{This VEV can be decomposed as follows:
\begin{align}
  \langle \hat{\Sigma} \rangle = \frac{3}{5}\hat{V}\cdot \mathrm{diag} (0,2,2,2,-3,-3) +  \frac{1}{5}\hat{V}\cdot \mathrm{diag} (5,-1,-1,-1,-1,-1) ~.
\end{align}
The first term breaks the SU(5) gauge group. It then follows that 
\begin{equation}
  \langle \Sigma \rangle = \frac{3}{5}\hat{V}\cdot \mathrm{diag} (2,2,2,-3,-3) ~, \qquad 
  \langle S \rangle = - \sqrt{\frac{12}{5}} \hat{V} ~.
\end{equation}
} This VEV spontaneously breaks the global SU(6) symmetry into $\mathrm{SU}(4) \otimes \mathrm{SU}(2) \otimes \mathrm{U}(1)$; the SU(5) gauge symmetry, which corresponds to the second to fifth rows/columns, is broken into the SM gauge group, $\mathrm{SU}(3)_C \otimes \mathrm{SU} (2)_L  \otimes \mathrm{U}(1)_Y$. The SU(3)$_C$ gauge group is inside the SU(4) global group. The symmetry breaking of $\mathrm{SU} (6) \to \mathrm{SU}(4) \otimes \mathrm{SU}(2) \otimes \mathrm{U}(1)$ yields $35 - (15+3+1) = 16$ NG bosons, among which 12 are absorbed by the massive gauge bosons corresponding to the broken generators of $\mathrm{SU}(5) \to \mathrm{SU}(3)_C \otimes \mathrm{SU} (2)_L  \otimes \mathrm{U}(1)_Y$. The rest four NG bosons, together with their SUSY partner fields, appear as physical NG chiral superfields---they are dubbed as the Mixed-type superfields in Ref.~\cite{Bando:1984cc}. As we see below, these four massless chiral superfields can be identified as the MSSM Higgs superfields.

We now calculate the mass spectrum of the component fields: 
\begin{align}
  H &= 
  \begin{pmatrix}
    H^1_C \\ H^2_C \\ H^3_C \\ H_u^+ \\ H_u^0 
  \end{pmatrix}
  ~, \qquad 
  \bar{H} =
  \begin{pmatrix}
    \bar{H}_{C1} \\ \bar{H}_{C2} \\ \bar{H}_{C3} \\ H_d^- \\ - H_d^0
  \end{pmatrix}
  ~, \\[3pt] 
  \Sigma & = 
  \begin{pmatrix}
    \Sigma_8&\Sigma_{(3,2)} \\
    \Sigma_{(3^*,2)} & \Sigma_3
   \end{pmatrix}
   +\frac{1}{2\sqrt{15}}
   \begin{pmatrix}
    2 \1_3 &0\\0&-3 \1_2
   \end{pmatrix}
   \Sigma_{24}~.
\end{align}
The adjoint Higgs fields $\Sigma_8$ and $\Sigma_3$ have the identical mass 
\begin{equation}
  M_\Sigma \equiv M_{\Sigma_8} = M_{\Sigma_3} = \frac{3}{2} \lambda \hat{V} ~.
  \label{eq:msigma}
\end{equation}
The components $\Sigma_{(3,2)} $ and $\Sigma_{(3^*,2)}$ are massless NG fields and absorbed by the SU(5) gauge fields to be massive. The component $\Sigma_{24}$ mixes with the SU(5) singlet field $S$, whose mass eigenvalues are found to be $3\lambda\hat{V}/2$ and $\lambda\hat{V}/2$. The color triplet Higgs fields $H_C$ and $\bar{H}_C$  acquire a mass of 
\begin{equation}
  M_{H_C} = \frac{3}{2} \lambda \hat{V} ~,
  \label{eq:mhc}
\end{equation}
which is equal to the adjoint Higgs mass. The MSSM Higgs mass is, on the other hand, computed as 
\begin{equation}
  M_H = \frac{1}{2} M - \frac{\lambda}{\sqrt{15}} \biggl(- \sqrt{\frac{12}{5}} \hat{V}\biggr) + \frac{\lambda}{2} \frac{3}{5}(-3)\hat{V}  = 0 ~,
  \label{eq:mh}
\end{equation}
verifying the expectation that $H_u$ and $H_d$ are the NG chiral superfields. We see that the doublet-triplet mass splitting is naturally realized in this setup. Finally, the mass of the SU(5) gauge bosons is found to be 
\begin{equation}
  M_X = 3 \sqrt{2} g_5 \hat{V} ~,
  \label{eq:mx}
\end{equation}
with $g_5$ the SU(5) gauge coupling constant. 

Notice that even though the SU(6) global symmetry is explicitly broken by the gauge interactions and $W_{\mathrm{Yukawa}}$, $H_u$ and $H_d$ remain massless in the SUSY limit because of the non-renormalization property of superpotential; in particular, radiative corrections to the K\"{a}hler potential, which give multiplicative wave-function renormalization factors to the component fields, do not generate a mass for $H_u$ and $H_d$ (see, \textit{e.g.}, Ref.~\cite{Anselm:1986um}). 

Next, we consider the effect of SUSY breaking. We here assume that the SUSY-breaking effect is mediated to the Higgs sector such that the global SU(6) symmetry is respected. The soft SUSY-breaking terms in the Higgs sector, then, have the form 
\begin{align}
  \mathcal{L}_{\mathrm{soft}} &= -\biggl( \frac{1}{3} A_\lambda \lambda \mathrm{Tr} \hat{\Sigma}^3 + \frac{1}{2} B_M M \mathrm{Tr} \hat{\Sigma}^2 + \mathrm{h.c.}\biggr) - 2 m_{\hat{\Sigma}}^2 \mathrm{Tr} (\hat{\Sigma}^\dagger \hat{\Sigma}) ~,
  \label{eq:lsoft}
\end{align}
where we use the same symbols for the scalar components of the Higgs fields. In terms of the SU(5) representations, these soft terms are expressed as 
\begin{align}
  \mathcal{L}_{\mathrm{soft}} = - &\biggl[\frac{1}{3} \lambda A_\lambda \mathrm{Tr} (\Sigma^3) + \frac{1}{2} \lambda A_\lambda \bar{H}\Sigma H
  - \frac{1}{3\sqrt{15}} \lambda A_\lambda S^3 -\frac{1}{\sqrt{15}} \lambda A_\lambda S \bar{H}H \nonumber \\ 
  &+ \frac{1}{\sqrt{60}} \lambda A_\lambda S \mathrm{Tr} (\Sigma^2) +\frac{1}{2}  B_M M \mathrm{Tr} (\Sigma^2) + \frac{1}{2} B_M M \bar{H}H
  + \frac{1}{4} B_M M S^2
  + \mathrm{h.c.}
  \biggr] \nonumber \\ 
  &- m_{\hat{\Sigma}}^2 \left[ |S|^2 
  %+ |H|^2 + |\bar{H}|^2 
  + H^\dagger H + \bar{H}^\dagger \bar{H}   
  + 2 \mathrm{Tr} (\Sigma^\dagger \Sigma) \right] ~.
\end{align}

In the presence of the soft SUSY-breaking terms, the VEV of $\hat{\Sigma}$ shifts from the one in Eq.~\eqref{eq:vev}~\cite{Hall:1983iz}. The $F$-term is also induced in $\langle \hat{\Sigma} \rangle$. We find 
\begin{equation}
  \langle \hat{\Sigma} \rangle = \left( \hat{V}  + \Delta \hat{V}  + F_{\hat{\Sigma}} \theta^2 \right)\cdot \mathrm{diag} (1,1,1,1,-2,-2) ~,
  \label{eq:vevwsoft}
\end{equation}
with 
\begin{align}
  \Delta \hat{V} &= \frac{2}{\lambda} (A_\lambda - B_M) -\frac{4}{\lambda^2 \hat{V}} \left( A_\lambda^2 - 3 A_\lambda B_M + 2 B_M^2 + m_{\hat{\Sigma}}^2  \right) + \mathcal{O}(M_{\mathrm{SUSY}}^3/M_{\mathrm{GUT}}^2)  ~,\\[2pt] 
  F_{\hat{\Sigma}} &= (A_\lambda - B_M) \hat{V} + \frac{2}{\lambda} \left( A_\lambda B_M - B_M^2 - m_{\hat{\Sigma}}^2 \right)
  + \mathcal{O} (M_{\mathrm{SUSY}}^3/M_{\mathrm{GUT}}) 
  ~,
\end{align}
where $M_{\mathrm{GUT}}$ and $M_{\mathrm{SUSY}}$ are the GUT and SUSY-breaking scales, and we take all of the parameters to be real just for simplicity. As we see in the next section, these terms generate the mass terms for the MSSM Higgs fields.

%%%%%%%%%%%%%%%%%%%%%%%%%%%%%%%%%%%%%%%%%%%%%%%%%%%%%%%%
\section{GUT-scale matching conditions}
\label{sec:matchingconditions}
%%%%%%%%%%%%%%%%%%%%%%%%%%%%%%%%%%%%%%%%%%%%%%%%%%%%%%%%

%%%%%%%%%%%%%%%%%%%%%%%%%%%%%%%%%%%%%%%
\subsection{Gauge coupling constants}
\label{sec:gcu}
%%%%%%%%%%%%%%%%%%%%%%%%%%%%%%%%%%%%%%%

The GUT-scale particles affect gauge coupling unification via threshold corrections. We can, therefore, extract the information on the GUT-scale mass spectrum from the mismatch of the gauge coupling constants at the GUT scale~\cite{Hisano:1992mh,Hisano:1992jj,Hisano:2013cqa}. At the scale $Q_G$ near the GUT scale, the one-loop matching conditions for
the gauge coupling constants in the $\overline{\rm DR}$ scheme \cite{Siegel:1979wq} are given by~\cite{Weinberg:1980wa, Hall:1980kf}
\begin{align}
 \frac{1}{g_1^2(Q_{G})}&=\frac{1}{g_5^2(Q_G)}
+\frac{1}{8\pi^2}\biggl[
\frac{2}{5}
\ln \frac{Q_{G}}{M_{H_C}}-10\ln\frac{Q_{G}}{M_X}
\biggr]~,\nonumber \\
 \frac{1}{g_2^2(Q_{G})}&=\frac{1}{g_5^2(Q_{G})}
+\frac{1}{8\pi^2}\biggl[
2\ln \frac{Q_{G}}{M_\Sigma}-6\ln\frac{Q_{G}}{M_X}
\biggr]~,\nonumber \\
 \frac{1}{g_3^2(Q_{G})}&=\frac{1}{g_5^2(Q_{G})}
+\frac{1}{8\pi^2}\biggl[
\ln \frac{Q_{G}}{M_{H_C}}+3\ln \frac{Q_{G}}
{M_\Sigma}-4\ln\frac{Q_{G}}{M_X}
\biggr]~,
\label{eq:gc_matching}
\end{align}
where $g_1 \equiv \sqrt{5/3}\, g'$, $g_2$, and $g_3$ are the gauge coupling constants of the SM gauge interactions. Now note that from Eq.~\eqref{eq:msigma} and Eq.~\eqref{eq:mhc}, we have $M_\Sigma = M_{H_C}$; we can, therefore, determine $g_5$, $M_X$, and $M_{H_C} = M_\Sigma$ using the above three equations by solving the RGEs of the gauge coupling constants up to the scale $Q_G$ for a given SUSY mass spectrum. From Eq.~\eqref{eq:gc_matching} with $M_\Sigma = M_{H_C}$, we obtain 
\begin{align}
  \alpha_5 (Q_G) &= 3 \biggl[ -\frac{10}{\alpha_1 (Q_G)} + \frac{24}{\alpha_2 (Q_G)} - \frac{11}{\alpha_3 (Q_G)} \biggr]^{-1} ~, \label{eq:alp5sol}\\[2pt] 
  M_X &= Q_G \exp \biggl[\frac{\pi}{6} \biggl(\frac{5}{\alpha_1 (Q_G)} - \frac{9}{\alpha_2 (Q_G)} + \frac{4}{\alpha_3 (Q_G)}\biggr)\biggr] ~,  \\[2pt] 
  M_{H_C} &= Q_G \exp \biggl[\frac{5\pi}{6} \biggl(- \frac{1}{\alpha_1 (Q_G)} + \frac{3}{\alpha_2 (Q_G)} - \frac{2}{\alpha_3 (Q_G)}\biggr)\biggr] ~, 
\end{align}
where $\alpha_a \equiv g_a^2/(4\pi)$ ($a = 1,2,3, 5$). From these quantities, we can determine the coupling $\lambda$ by using Eq.~\eqref{eq:mhc} and Eq.~\eqref{eq:mx}: 
\begin{equation}
  \lambda = 2\sqrt{2} g_5 \frac{M_{H_C}}{M_X} ~.
  \label{eq:lamsol}
\end{equation}

%%%%%%%%%%%%%%%%%%%%%%%%%%%%%%%%%%%%%%%%%%%%%%%%%%%%%%%%%%%%%%%%
\begin{figure}
  \centering
  \subcaptionbox{\label{fig:newlambda}
  $\lambda$
  }
  {\includegraphics[width=0.495\textwidth]{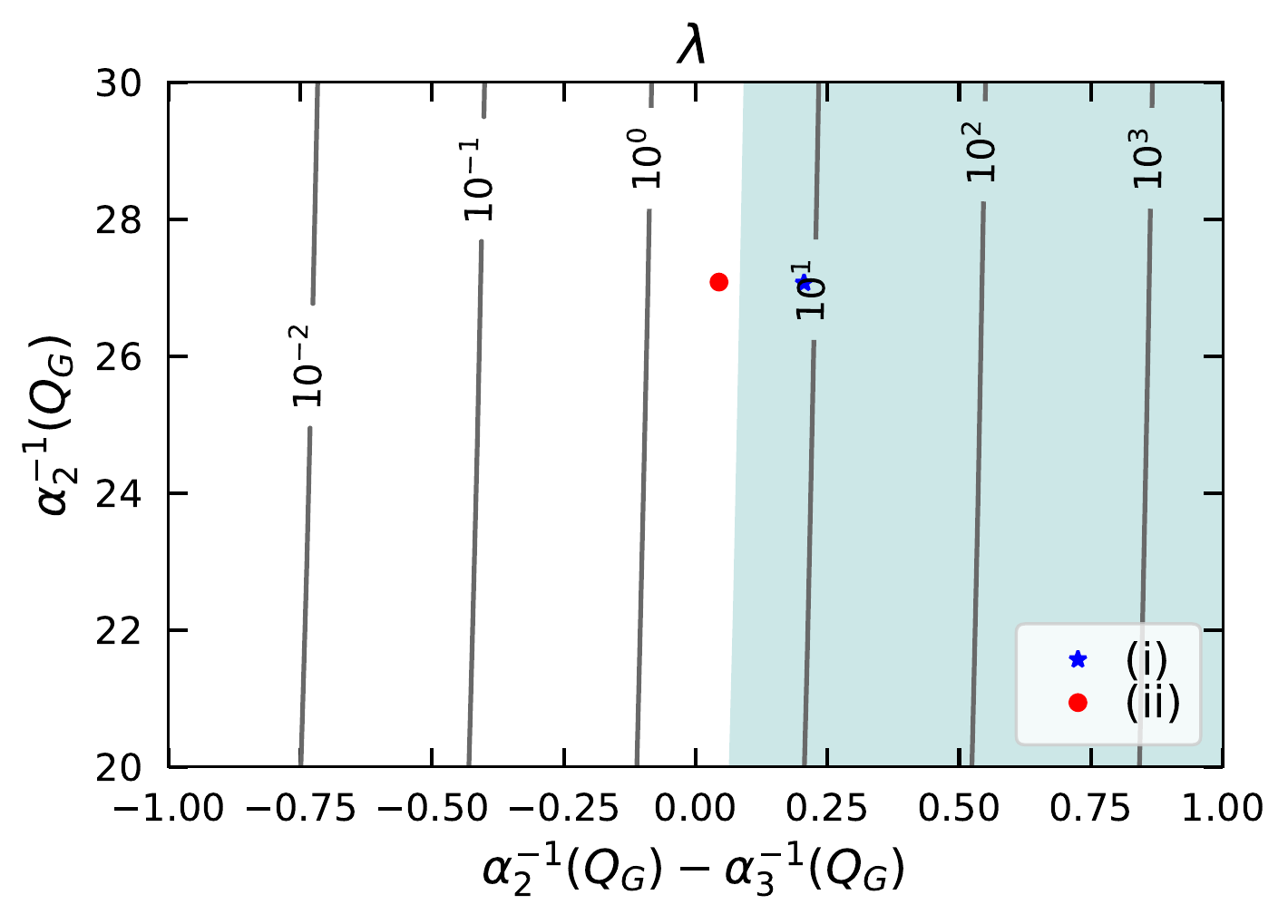}}
  \vspace{15pt}
  \subcaptionbox{\label{fig:newalpha_5}
  $\alpha_5^{-1} (Q_G)$
  }
  {\includegraphics[width=0.495\textwidth]{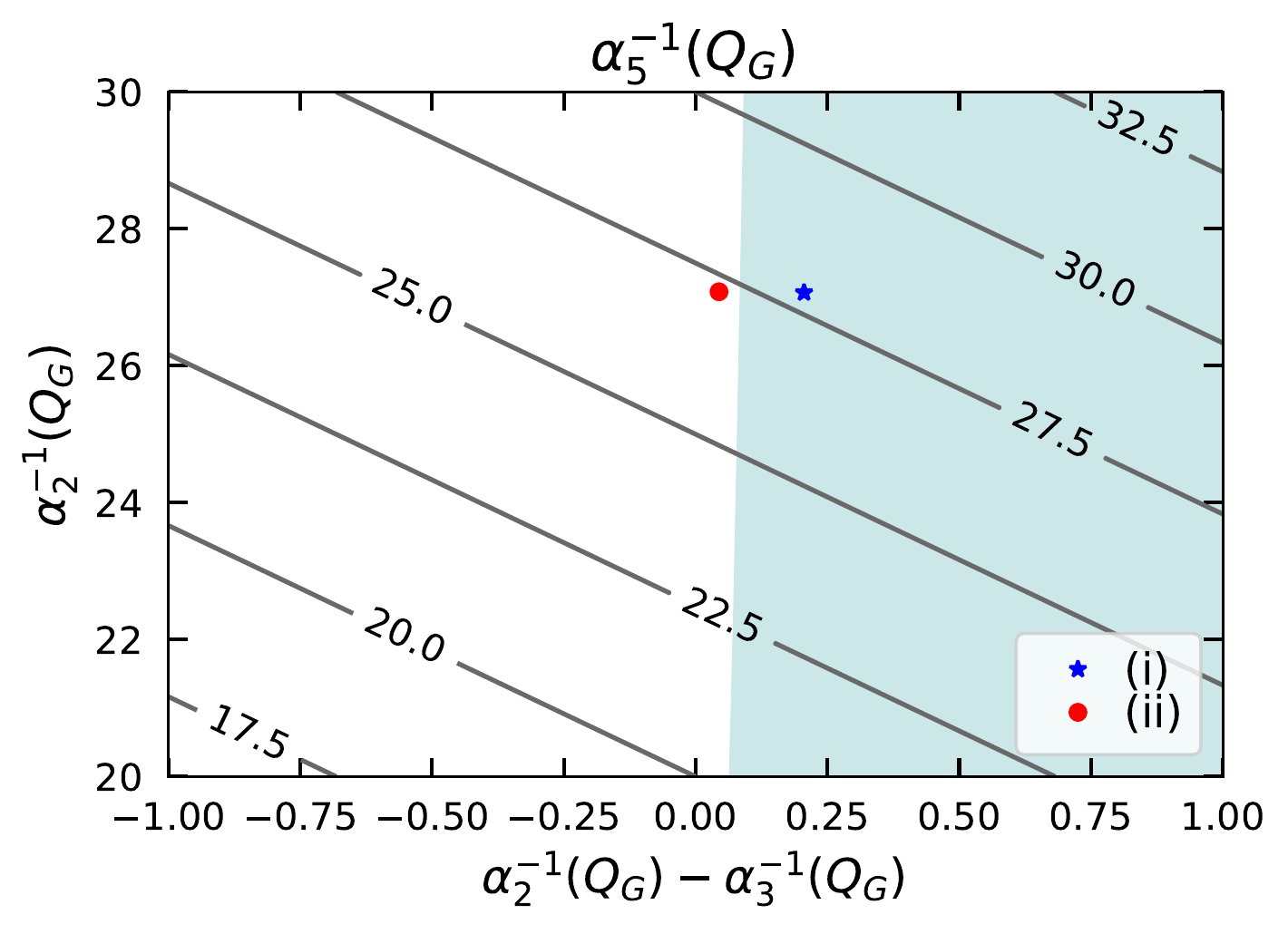}}
  \subcaptionbox{\label{fig:newMHX}
  $M_{H_C}/M_X$
  }
  {\includegraphics[width=0.495\textwidth]{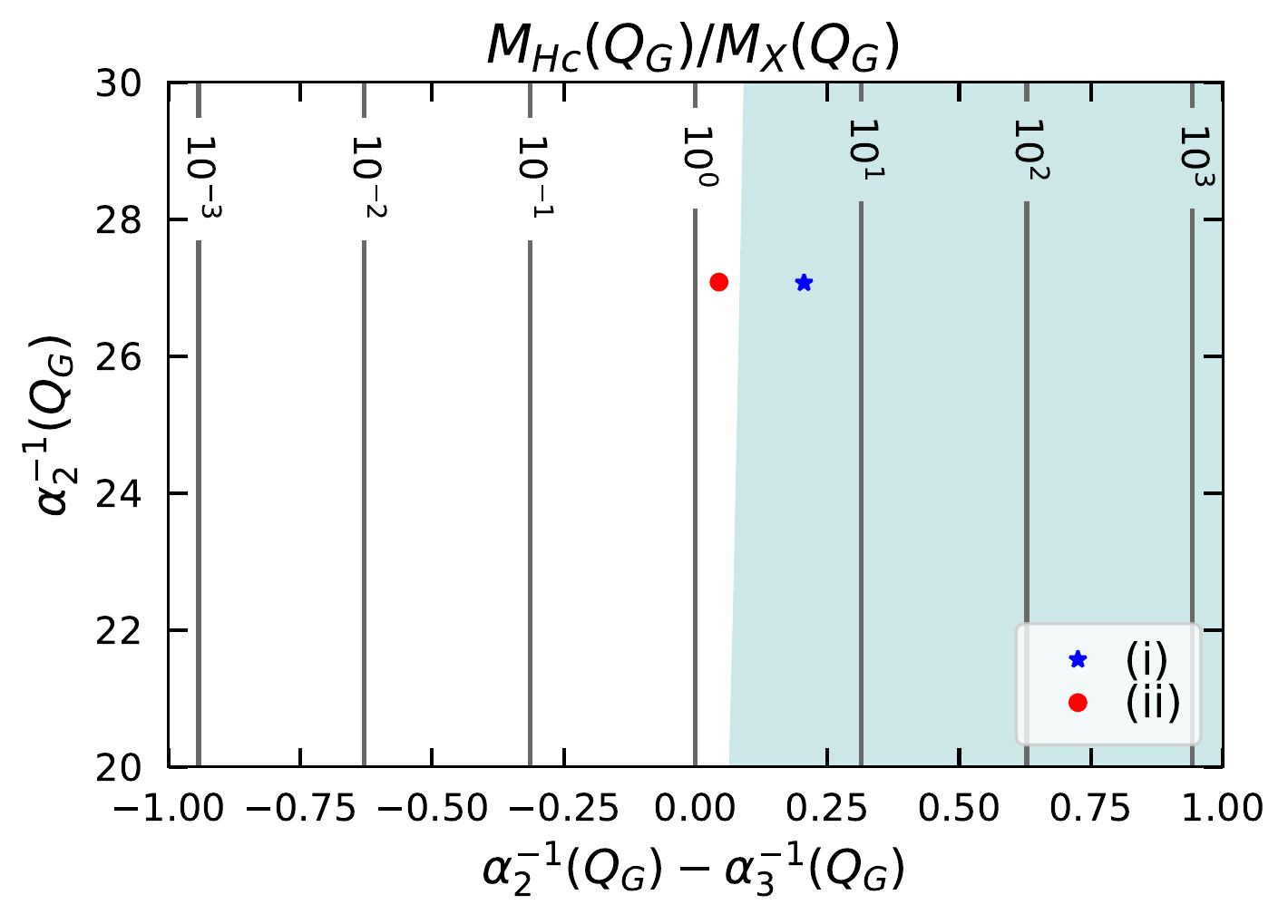}}
  \caption{
    Contour plots of (a) $\lambda$, (b) $\alpha_5^{-1} (Q_G)$, and (c) $M_{H_C}/M_X$ in the $\left(\alpha_2^{-1} (Q_G) - \alpha_3^{-1} (Q_G) \right)$ -- $\alpha_2^{-1} (Q_G)$ plane
    in the minimal NG Higgs SUSY SU(5) GUT model. The green shaded region represents $\lambda > \sqrt{4\pi}$, which is ruled out by the perturbativity condition. The blue star and red dot correspond to the model points in the constrained MSSM discussed in Sec.~\ref{sec:cmssm}.
  }
\label{fig:alp23}
\end{figure}
%%%%%%%%%%%%%%%%%%%%%%%%%%%%%%%%%%%%%%%%%%%%%%%%%%%%%%%%%%%%%%%%

Now let us choose the scale $Q_G$ such that $g_1 (Q_G) = g_2 (Q_G)$. In this case, we have the relations 
\begin{align}
  \alpha_5 (Q_G) &= 3 \biggl[\frac{14}{\alpha_2 (Q_G)} - \frac{11}{\alpha_3 (Q_G)} \biggr]^{-1} ~, \\[2pt]
  \frac{M_{H_C}}{M_X} &= \exp \biggl[\frac{7\pi}{3} \biggl(\frac{1}{\alpha_2 (Q_G)} - \frac{1}{\alpha_3 (Q_G)} \biggr)\biggr] ~.
  \label{eq:mHX}
\end{align}
The above equations indicate that $\lambda$, $M_{H_C}/M_X$, and $\alpha_5$ are given as functions of $\alpha_2 (Q_G)$ and $\alpha_3(Q_G)$. 

In Fig.~\ref{fig:alp23}, we show contour plots for these values in the $(\alpha_2^{-1} (Q_G) - \alpha_3^{-1} (Q_G) )$ -- $\alpha_2^{-1} (Q_G)$ plane. As we see from Fig.~\ref{fig:newlambda}, $\lambda$ increases as $(\alpha_2^{-1} (Q_G) - \alpha_3^{-1} (Q_G) )$ gets larger, with little dependence on $\alpha_2^{-1} (Q_G)$, and becomes non-perturbative for $(\alpha_2^{-1} (Q_G) - \alpha_3^{-1} (Q_G) ) \gtrsim 0$; the region where $\lambda > \sqrt{4\pi}$ is represented by the green shade in Fig.~\ref{fig:alp23}. We thus conclude that $(\alpha_2^{-1} (Q_G) - \alpha_3^{-1} (Q_G) ) \lesssim 0$, \textit{i.e.}, $\alpha_2(Q_G) \gtrsim \alpha_3 (Q_G)$ is favored in the present setup, which imposes a non-trivial constraint on the low-energy SUSY spectrum. On the other hand, Fig.~\ref{fig:newalpha_5} shows that $\alpha_5 (Q_G)$ remains perturbative over the parameter region shown in this figure.

As can be seen from Eq.~\eqref{eq:mHX} and Fig.~\ref{fig:newMHX}, $M_{H_C}/M_X$ is a function of $(\alpha_2^{-1} (Q_G) - \alpha_3^{-1} (Q_G) )$ only, and $M_{H_C} < M_X$ ($M_{H_C} > M_X$) for $\alpha_2 (Q_G) > \alpha_3 (Q_G)$ ($\alpha_2 (Q_G) < \alpha_3 (Q_G)$). As discussed above, the perturbativity condition on $\lambda$ favors $\alpha_2(Q_G) \gtrsim \alpha_3 (Q_G)$, and hence $M_{H_C} \lesssim M_X$. On the other hand, if $\alpha_2 (Q_G)$ is considerably larger than $\alpha_3 (Q_G)$, then $M_{H_C}$ becomes much smaller than $M_X$. In this case, the proton-decay rate induced by the exchange of the color-triplet Higgs multiplet is enhanced, making it difficult to avoid the experimental bound. We, therefore, expect that the proton-decay limit gives a lower limit on $\alpha_2^{-1} (Q_G) - \alpha_3^{-1} (Q_G) $, which further restricts the low-energy SUSY mass spectrum.

To see qualitatively the dependence of $\alpha_2^{-1} (Q_G) - \alpha_3^{-1} (Q_G) $ on the low-energy SUSY spectrum, we use the one-loop RGEs for the gauge coupling constants to obtain analytical expressions for $\alpha_a (Q_G)$ $(a = 1,2,3)$. We have 
\begin{align}
  \frac{1}{\alpha_1 (Q_G)} &= \frac{1}{\alpha_1 (m_Z)} - \frac{1}{2\pi} \biggl\{ \frac{41}{10} \ln \biggl(\frac{Q_G}{m_Z}\biggr) + \frac{1}{10} \ln \biggl(\frac{Q_G}{m_A} \biggr) + \frac{2}{5} \ln \biggl(\frac{Q_G}{m_{\widetilde{H}}}\biggr)\nonumber \\[2pt] 
  &+ \frac{1}{5} \sum_i \biggl[
  \frac{1}{6} \ln \biggl(\frac{Q_G}{m_{\widetilde{Q}_i}}\biggr)  
  +\frac{4}{3} \ln \biggl(\frac{Q_G}{m_{\widetilde{\bar{u}}_i}}\biggr)  
  + \frac{1}{3} \ln \biggl(\frac{Q_G}{m_{\widetilde{\bar{d}}_i}}\biggr) \nonumber \\[2pt]
  &+ \frac{1}{2} \ln \biggl(\frac{Q_G}{m_{\widetilde{L}_i}}\biggr)  
  + \ln \biggl(\frac{Q_G}{m_{\widetilde{\bar{e}}_i}}\biggr)  
  \biggr] \biggr\}~, \\[3pt]
  \frac{1}{\alpha_2 (Q_G)} &= \frac{1}{\alpha_2 (m_Z)} - \frac{1}{2\pi} 
  \biggl\{
  -\frac{19}{6} \ln\biggl(\frac{Q_G}{m_Z}\biggr)  
  + \frac{4}{3} \ln \biggl(\frac{Q_G}{m_{\widetilde{W}}}\biggr)
  + \frac{1}{6} \ln\biggl(\frac{Q_G}{m_A}\biggr)  \nonumber \\[2pt]
  &+ \frac{2}{3} \ln \biggl(\frac{Q_G}{m_{\widetilde{H}}}\biggr)
  + \frac{1}{6} \sum_i \biggl[ 3\ln \biggl(\frac{Q_G}{m_{\widetilde{Q}_i}}\biggr)
  + \ln \biggl(\frac{Q_G}{m_{\widetilde{L}_i}}\biggr)
  \biggr]
  \biggr\} ~, \\[3pt]
  \frac{1}{\alpha_3 (Q_G)} &= \frac{1}{\alpha_3 (m_Z)} - \frac{1}{2\pi} 
  \biggl\{-7 \ln\biggl(\frac{Q_G}{m_Z}\biggr)  
  + 2 \ln \biggl(\frac{Q_G}{m_{\widetilde{g}}}\biggr) \nonumber \\[2pt]
  & + \frac{1}{6} \sum_i \biggl[ 2\ln \biggl(\frac{Q_G}{m_{\widetilde{Q}_i}}\biggr)
  + \ln \biggl(\frac{Q_G}{m_{\widetilde{\bar{u}}_i}}\biggr)
  + \ln \biggl(\frac{Q_G}{m_{\widetilde{\bar{d}}_i}}\biggr)
  \biggr]
  \biggr\}~,
\end{align}
where $m_Z$, $m_A$, $m_{\widetilde{H}}$, $m_{\widetilde{W}}$, $m_{\widetilde{g}}$, $m_{\widetilde{Q}_i}$, $m_{\widetilde{L}_i}$, $m_{\widetilde{\bar{u}}_i}$, $m_{\widetilde{\bar{d}}_i}$, $m_{\widetilde{\bar{e}}_i}$ are the masses of $Z$ boson, the heavy Higgs doublet, higgsino, wino, gluino, and sfermions, respectively, and $i = 1,2,3$ is the generation index. From the condition $\alpha_1 (Q_G) = \alpha_2 (Q_G)$, we obtain the unification scale $Q_G$ as 
\begin{align}
  Q_G &= m_Z \cdot \exp \biggl[\frac{5\pi}{14} \biggl\{\frac{1}{\alpha_1 (m_Z)} - \frac{1}{\alpha_2 (m_Z)} \biggr\} \biggr]
  \nonumber \\ & \times 
  \biggl(\frac{m_Z^{25}}{m_{\widetilde{W}}^{20}  \cdot m_{\widetilde{H}}^{4} \cdot  m_A} \biggr)^{\frac{1}{84}} \prod_i \biggl[ \biggl( 
    \frac{m_{\widetilde{\bar{u}}_i}^{4} \cdot m_{\widetilde{\bar{e}}_i}^{3}}{m_{\widetilde{Q}_i}^{7}}
    \biggr)^{\frac{1}{84}} \biggl( 
    \frac{m_{\widetilde{\bar{d}}_i}}{m_{\widetilde{L}_i}}
    \biggr)^{\frac{1}{84}}\biggr]~. 
    \label{eq:Q_G}
\end{align}
By using this, we then have 
\begin{align}
  \frac{1}{\alpha_2 (Q_G)} - \frac{1}{\alpha_3 (Q_G)} 
  =& - \frac{5}{7} \frac{1}{\alpha_1 (m_Z)} + \frac{12}{7} \frac{1}{\alpha_2 (m_Z)} - \frac{1}{\alpha_3 (m_Z)} \nonumber \\[2pt]
  &+ \frac{1}{28\pi}\ln\biggl(
  \frac{m_{\widetilde{W}}^{32}  \cdot m_{\widetilde{H}}^{12} \cdot m_A^3}{m_Z^{19} \cdot m_{\widetilde{g}}^{28}} \biggr)
  + \frac{1}{28\pi} \sum_i \ln\biggl[ \biggl( 
  \frac{m_{\widetilde{Q}_i}^7}{m_{\widetilde{\bar{u}}_i}^5 \cdot m_{\widetilde{\bar{e}}_i}^2}
  \biggr) \biggl( 
  \frac{m_{\widetilde{L}_i}^3}{m_{\widetilde{\bar{d}}_i}^3}
  \biggr)\biggr]~.
  \label{eq:alpha2alpha3}
\end{align}
Note that $\widetilde{Q}_i$, $\widetilde{\bar{u}}_i$, and $\widetilde{\bar{e}}_i$ ($\widetilde{L}_i$ and $\widetilde{\bar{d}}_i$) come from the same $\mathbf{10}_i$ ($\bar{\mathbf{5}}_i$) multiplet in SU(5). As can be seen in the last terms in Eq.~\eqref{eq:Q_G} and Eq.~\eqref{eq:alpha2alpha3}, these sfermions give no contribution if the masses of the components in the same SU(5) multiplet are equal. We also see from Eq.~\eqref{eq:Q_G} that the scale $Q_G$ decreases as $m_{\widetilde{W}}$, $m_{\widetilde{H}}$, and $m_A$ increase, with the strongest dependence on the wino mass. On the other hand, Eq.~\eqref{eq:alpha2alpha3} shows that $\alpha_2^{-1} (Q_G) - \alpha_3^{-1} (Q_G) $ increases as these masses increase, whilst it decreases as the gluino mass increases. As shown in Fig.~\ref{fig:alp23}, the perturbativity condition sets an upper limit on $\alpha_2^{-1} (Q_G) - \alpha_3^{-1} (Q_G) $; this then leads to upper limits on $m_{\widetilde{W}}$, $m_{\widetilde{H}}$, and $m_A$, and a lower limit on $m_{\widetilde{g}}$. 

Notice that the above analysis based on 1-loop RGEs is insufficient since the two-loop RGE effect is as large as that of one-loop threshold corrections. We perform a numerical computation to include the two-loop RGE effect in the subsequent section.\footnote{See Appendix~\ref{sec:RG} for the formulae used in this analysis. }

%%%%%%%%%%%%%%%%%%%%%%%%%%%%%%%%%%%%%%%%%
\subsection{Higgs mass parameters}
%%%%%%%%%%%%%%%%%%%%%%%%%%%%%%%%%%%%%%%%%

Next, we discuss the matching conditions for the MSSM Higgs mass terms. As seen in Eq.~\eqref{eq:mh}, the $\mu$-term of $H_u$ and $H_d$ vanishes in the SUSY limit. Once the SUSY-breaking effect is included, the shift in $\langle \hat{\Sigma} \rangle$ generates an effective $\mu$-term, which is given by 
\begin{equation}
  \mu =  - (A_\lambda - B_M) ~.
  \label{eq:mu_match}
\end{equation}
This shows that the size of the $\mu$-term becomes $\mathcal{O} (M_{\mathrm{SUSY}})$ automatically. In particular, there is no ``$\mu$ problem'' in this model. The soft SUSY-breaking terms for $H_u$ and $H_d$ are 
\begin{align}
  \mathcal{L}_{\mathrm{soft}} = - (B \mu H_u H_d + \mathrm{h.c.}) - m_{H_u}^2 |H_u|^2 - m_{H_d}^2 |H_d|^2 ~.
\end{align}
The matching condition for the $B$ term is calculated as 
\begin{align}
  B \mu &= \frac{1}{2} \lambda F_{\hat{\Sigma}} - \frac{1}{2} \lambda A_\lambda (\hat{V} + \Delta \hat{V}) + \frac{1}{2} \lambda B_M \hat{V} \nonumber \\ 
  &= - (A_\lambda - B_M)^2 - m_{\hat{\Sigma}}^2 ~.
  \label{eq:b_match}
\end{align}
Notice that the $\mathcal{O} (M_{\mathrm{SUSY}} M_{\mathrm{GUT}})$ terms cancel and therefore $B \mu$ is $\mathcal{O} (M_{\mathrm{SUSY}}^2)$. As shown in Ref.~\cite{Kawamura:1994ys} this cancellation is stable against renormalization. The soft mass terms are given by 
\begin{equation}
  m_{H_u}^2 = m_{H_d}^2 = m_{\hat{\Sigma}}^2~.
  \label{eq:mhu_match}
\end{equation}
It is interesting to note that the determinant of the Higgs mass matrix, which is the order parameter of electroweak symmetry breaking, vanishes at the GUT scale; from Eq.~\eqref{eq:mu_match}, Eq.~\eqref{eq:b_match}, and Eq.~\eqref{eq:mhu_match}, we find 
\begin{align}
  \left( m_{H_u}^2 + |\mu|^2 \right) \left( m_{H_d}^2 + |\mu|^2 \right)
  - |B\mu|^2 = 0 ~.
  \label{eq:higgsdet}
\end{align}
This is because we have assumed that the soft SUSY-breaking terms in the Higgs sector~\eqref{eq:lsoft} respect the global SU(6) symmetry and therefore we still have four massless NG fields, forming an SU(2)$_L$ doublet, even after the SUSY-breaking effect is included. The relation~\eqref{eq:higgsdet} is, however, violated by radiative corrections due to the presence of the SU(6)-breaking interactions, which results in radiative electroweak symmetry breaking at low energies~\cite{Ibanez:1982fr, Inoue:1982pi, Ibanez:1982ee, Ellis:1982wr, Ellis:1983bp, Alvarez-Gaume:1983drc}. 

Depending on the mediation mechanism, it is also possible that the SUSY-breaking effect is transmitted in an SU(6)-violating manner. In this case, the $\mu$ and $B\mu$ terms may receive additional contributions of $\mathcal{O}(M_{\mathrm{SUSY}})$ and $\mathcal{O}(M_{\mathrm{SUSY}}^2)$, respectively, which can violate the relation~\eqref{eq:higgsdet} by  $\mathcal{O}(M_{\mathrm{SUSY}}^4)$. Considering this, we do not strictly require the condition~\eqref{eq:higgsdet} and regard $\mu$ and $B \mu$ as free parameters in the following analysis.

%%%%%%%%%%%%%%%%%%%%%%%%%%%%%%%%%%%%%%%%%%%%%%%%%%%%%%%%
\section{Results}
\label{sec:results}
%%%%%%%%%%%%%%%%%%%%%%%%%%%%%%%%%%%%%%%%%%%%%%%%%%%%%%%%

We now calculate the masses of the GUT-scale particles, $M_X$ and $M_{H_C}$, for two SUSY-breaking scenarios. We first consider in Sec.~\ref{sec:cmssm} the CMSSM, where the soft SUSY-breaking masses of sfermions and gauginos are taken to be universal at the GUT scale. This setup is often regarded as a benchmark scenario of SUSY models, and we use this to show typical values of $\alpha_5$, $M_X$, and $M_{H_C}$. It is, however, found that this case is actually ruled out by the limit on the $p \to K^+ \bar{\nu}$ lifetime for the minimal NG Higgs SUSY SU(5). This limit can be evaded if squarks and sleptons lie in the PeV range; motivated by this, we consider high-scale SUSY-breaking scenarios in Sec.~\ref{sec:highscale} and show the predicted GUT-scale mass spectrum. We also calculate the lifetimes\footnote{The calculation of the partial decay widths of $p\to K^+ \bar{\nu}$ and $p \to \pi^0 e^+$ is reviewed in Appendix~\ref{sec:dim5} and \ref{sec:dim6}, respectively. } of $p\to K^+ \bar{\nu}$ and $p \to \pi^0 e^+$ and confront them with the current bounds and future prospects.\footnote{For a recent review on proton-decay searches, see Ref.~\cite{Dev:2022jbf}.  }

%%%%%%%%%%%%%%%%%%%%%%%%%%%%%%%%%%%
\subsection{Constrained MSSM}
\label{sec:cmssm}
%%%%%%%%%%%%%%%%%%%%%%%%%%%%%%%%%%%

The CMSSM is specified by five input parameters, $m_0$, $m_{1/2}$, $A_0$, $\tan \beta$, and $\mathrm{sgn} (\mu)$. The soft masses, gaugino masses, and the $A$-terms are set to be $m_0$, $m_{1/2}$, and $A_0$, respectively, at the GUT scale. $\tan \beta \equiv \langle H_u^0\rangle/ \langle H_d^0 \rangle$ is the ratio between the MSSM Higgs VEVs. The electroweak-symmetry breaking conditions determine the magnitude of $\mu$, but its sign is undetermined, so is an input parameter. The soft SUSY-breaking terms at low energies are obtained by solving RGEs, which then determine the SUSY spectrum. Despite the limited number of input parameters, the CMSSM is known to provide desirable SUSY spectra which can explain the observed values of the SM-like Higgs boson mass, $m_h = 125.25(17)~\mathrm{GeV}$~\cite{ParticleDataGroup:2020ssz}, and the dark matter abundance, $\Omega_{\mathrm{DM}} h^2 \simeq 0.12$~\cite{Planck:2018vyg}, without conflicting with the current experimental limits (see, \textit{e.g.}, Refs.~\cite{Ellis:2015rya, Ellis:2016tjc, GAMBIT:2017snp, Ellis:2018jyl, Bagnaschi:2018igf, Ellis:2019fwf, Wang:2021bcx, Evans:2021hyx} for recent studies).

To show typical GUT-scale spectra for the CMSSM, we consider the following two sets of parameters taken from the benchmark points presented in Ref.~\cite{Ellis:2019fwf}:
\begin{itemize}
  \item[(i)] $m_0 = 14.1~\mathrm{TeV}$, $m_{1/2} = 9.8~\mathrm{TeV}$, $A_0 = -3m_0$, $\tan \beta = 5$, $\mu > 0$.\footnote{The opposite sign of $A_0$ with respect to that adopted in Ref.~\cite{Ellis:2019fwf} is due to convention~\cite{Buchmueller:2010ai}. } 
  \item[(ii)] $m_0 = 27.9~\mathrm{TeV}$, $m_{1/2} = 9.5~\mathrm{TeV}$, $A_0 = 0$, $\tan \beta = 4$, $\mu > 0$.  
\end{itemize}
We use \texttt{SOFTSUSY4.1.12}~\cite{Allanach:2001kg, Allanach:2014nba} to compute the SUSY spectra and the gauge coupling constants at the GUT scale $Q_G$, which is defined by the condition $g_1(Q_G) = g_2 (Q_G)$. By using Eqs.~(\ref{eq:alp5sol}--\ref{eq:lamsol}), we then find 
\begin{align}
  \alpha_5 &= 0.036 ~, \quad 
  \lambda = 8.7 ~, \quad  
  M_X = 4.8 \times 10^{15}~\mathrm{GeV} ~, \quad 
  M_{H_C} = 2.2 \times 10^{16}~\mathrm{GeV} ~, 
\end{align}
for point (i) and 
\begin{align}
  \alpha_5 &= 0.037 ~, \quad 
  \lambda = 2.7 ~, \quad  
  M_X = 7.6 \times 10^{15}~\mathrm{GeV} ~, \quad 
  M_{H_C} = 1.1 \times 10^{16}~\mathrm{GeV} ~, 
\end{align}
for point (ii). We also show the predictions of these two benchmark points by the blue star and red dot in Fig.~\ref{fig:alp23}. In both of these cases, $\lambda$ is predicted to be quite large; in particular, for case (i), $\lambda$ is almost non-perturbative, so our analysis based on one-loop threshold corrections may not be valid. These results suggest that the requirement of the perturbativity of the $\lambda$ coupling gives an important constraint on the SUSY mass spectrum. 

In any case, as we mentioned at the beginning of this section, the minimal NG Higgs SUSY SU(5) with the CMSSM is incompatible with the $p \to K^+ \bar{\nu}$ bound from the Super-Kamiokande experiment~\cite{Super-Kamiokande:2014otb, Takhistov:2016eqm} due to relatively light SUSY spectra and the restrictive relation among the GUT-scale particle masses. As shown in Refs.~\cite{Ellis:2019fwf, Evans:2021hyx}, in the CMSSM, the observed values of the SM-like Higgs-boson mass and the dark matter abundance can be reproduced with an $\mathcal{O} (10)$~TeV SUSY-breaking scale. To evade the $p \to K^+ \bar{\nu}$ limit in this case, we need a relatively large color-triplet Higgs mass, $M_{H_C} \gtrsim 10^{17}~\mathrm{GeV}$. In the minimal SUSY SU(5), $M_{H_C}$ is related to other GUT-parameters as 
\begin{equation}
  M_{H_C} = \lambda_H  \frac{\left( M_X^2 M_\Sigma \right)^{\frac{1}{3}}}{\left( g_5^2 \lambda_\Sigma \right)^{\frac{1}{3}}} ~,
\end{equation}
where $\lambda_\Sigma$ and $\lambda_H$ correspond to the couplings for the first and second terms in Eq.~\eqref{eq:whiggs}, respectively. In the minimal SUSY SU(5), these two couplings can be different, whilst in the minimal NG Higgs GUT model they are related as 
\begin{equation}
  \lambda_\Sigma = 2 \lambda_H = \lambda ~.
  \label{eq:lamrel}
\end{equation}
The combination $\left( M_X^2 M_\Sigma \right)^{\frac{1}{3}}$ can be determined via the GUT-scale threshold corrections~\cite{Hisano:1992mh,Hisano:1992jj,Hisano:2013cqa, Pokorski:2017ueo, Pokorski:2019ete, Babu:2020ncc} as we have done in Sec.~\ref{sec:gcu}. It is found that the dependence of this quantity on low-energy SUSY mass spectra is small, with $\left( M_X^2 M_\Sigma \right)^{\frac{1}{3}} \simeq 10^{16}~\mathrm{GeV}$ (see, for instance, Refs.~\cite{Pokorski:2017ueo, Pokorski:2019ete, Babu:2020ncc}). To obtain $M_{H_C} \gtrsim 10^{17}~\mathrm{GeV}$, therefore, we need $\lambda_\Sigma \ll \lambda_H$; however, this is incompatible with the relation~\eqref{eq:lamrel} in the minimal NG Higgs GUT model. In other words, for a perturbative value of $\lambda$, the color-triplet Higgs mass is $\lesssim 10^{16}~\mathrm{GeV}$ in the minimal NG Higgs GUT model, and this is too low to evade the proton-decay limit if SUSY particles lie around $\mathcal{O}(10)$~TeV. We thus do not explore the CMSSM in more detail and instead consider a high-scale SUSY scenario in what follows.

%%%%%%%%%%%%%%%%%%%%%%%%%%%%%%%
\subsection{High-scale SUSY}
\label{sec:highscale}
%%%%%%%%%%%%%%%%%%%%%%%%%%%%%%%

SUSY models with a SUSY-breaking scale of $\mathcal{O}(1)$~PeV have attracted wide attention~\cite{Wells:2003tf, Wells:2004di, Arkani-Hamed:2004ymt, Giudice:2004tc, Arkani-Hamed:2004zhs, Arkani-Hamed:2005zuc, Hall:2011jd, Hall:2012zp, Ibe:2011aa, Ibe:2012hu, Arvanitaki:2012ps, Arkani-Hamed:2012fhg, Evans:2013lpa, Evans:2013dza} especially after the discovery of the Higgs boson~\cite{ATLAS:2012yve, CMS:2012qbp}, since in these models the observed value of its mass can easily be explained by means of large threshold corrections by heavy stops~\cite{Okada:1990vk, Okada:1990gg, Ellis:1990nz, Haber:1990aw, Ellis:1991zd}. In addition, the PeV-scale SUSY scenario has several attractive features from a phenomenological viewpoint; i) the minimal SUSY SU(5) GUT with low-scale SUSY-breaking suffers from the rapid proton-decay problem~\cite{Goto:1998qg, Murayama:2001ur}, which can be evaded with PeV-scale sfermions~\cite{Hisano:2013exa, Nagata:2013sba, Evans:2015bxa, Evans:2019oyw}; ii) the SUSY flavor/CP problems are alleviated~\cite{Gabbiani:1996hi, Moroi:2013sfa, McKeen:2013dma, Altmannshofer:2013lfa, Fuyuto:2013gla}; iii) the cosmological gravitino problem~\cite{Weinberg:1982zq, Khlopov:1984pf, Ellis:1984eq, Lindley:1984bg, Ellis:1984er, Juszkiewicz:1985gg, Ellis:1990nb, Kawasaki:1994af, Moroi:1995fs, Holtmann:1998gd, Kawasaki:2000qr, Cyburt:2002uv, Kawasaki:2004yh, Kohri:2005wn, Ellis:2005ii, Kawasaki:2008qe, Cyburt:2010vz, Kawasaki:2017bqm} is highly relaxed. In particular, the feature i) is desirable for the minimal NG Higgs GUT, given that the proton lifetime limit is problematic in the case of the CMSSM as we have seen in Sec.~\ref{sec:cmssm}.

A concrete realization of the high-scale SUSY scenario is provided by the assumption that there is no gauge-singlet SUSY-breaking field in the hidden sector. In this case, the soft masses of SUSY particles are induced via gravity mediation and their size is $\mathcal{O} (m_{3/2})$, where $m_{3/2}$ is the gravitino mass. The gaugino masses are, on the other hand, generated only radiatively and thus suppressed by a loop factor compared with $m_{3/2}$. A famous mechanism for such quantum effect is anomaly mediation~\cite{Randall:1998uk, Giudice:1998xp}. In addition to this, threshold corrections by heavy Higgs fields~\cite{Pierce:1996zz}, vector-like matter fields~\cite{Pomarol:1999ie, Nelson:2002sa,  Hsieh:2006ig,Gupta:2012gu, Nakayama:2013uta, Harigaya:2013asa, Evans:2014xpa}, or GUT-scale fields~\cite{Evans:2019oyw} give rise to gaugino masses. 
The higgsino mass parameter, $\mu$, is given by Eq.~\eqref{eq:mu_match} in the current model, and in the absence of the singlet SUSY-breaking field, $A_\lambda = 0$ at the classical level, whilst the $B$-term can be $\mathcal{O} (m_{3/2})$ because of the Giudice-Masiero mechanism~\cite{Giudice:1988yz, Inoue:1991rk}---we, therefore, expect $\mu = \mathcal{O} (m_{3/2})$.

We perform an analysis similar to that in the previous subsection for this type of mass spectrum. In the present case, there is a large mass hierarchy between the gauginos and other SUSY particles. We, therefore, take an effective-theoretical approach, rather than just using \texttt{SOFTSUSY}, in order to avoid large logarithmic corrections. Between the GUT and electroweak scales, we introduce two threshold mass scales---the gaugino and SUSY scales. Below the gaugino mass scale, $Q_{\mathrm{gaugino}}$, the theory is just the SM, while above the SUSY scale, $Q_{\mathrm{SUSY}}$, the theory is the MSSM. Between $Q_{\mathrm{gaugino}}$ and $Q_{\mathrm{SUSY}}$, the effective theory consists of the SM plus gauginos. We use two-loop RGEs for the gauge coupling constants and one-loop RGEs for the Yukawa couplings,\footnote{Since the Yukawa couplings enter into the gauge coupling RGEs at the two-loop level, the one-loop RGEs are sufficient for the Yukawa couplings. } which we summarize in Appendix~\ref{sec:RGEs}. The values of input parameters and the expressions for matching conditions are given in Sec.~\ref{sec:input} and Sec.~\ref{sec:thcorr}, respectively.

%%%%%%%%%%%%%%%%%%%%%%%%%%%%%%%%%%%%%%%%%%%%%%%%%%%%%%%%%%%%%%%%
\begin{figure}
  \centering
  \subcaptionbox{\label{fig:lambda}
  $\lambda$
  }
  {\includegraphics[width=0.495\textwidth]{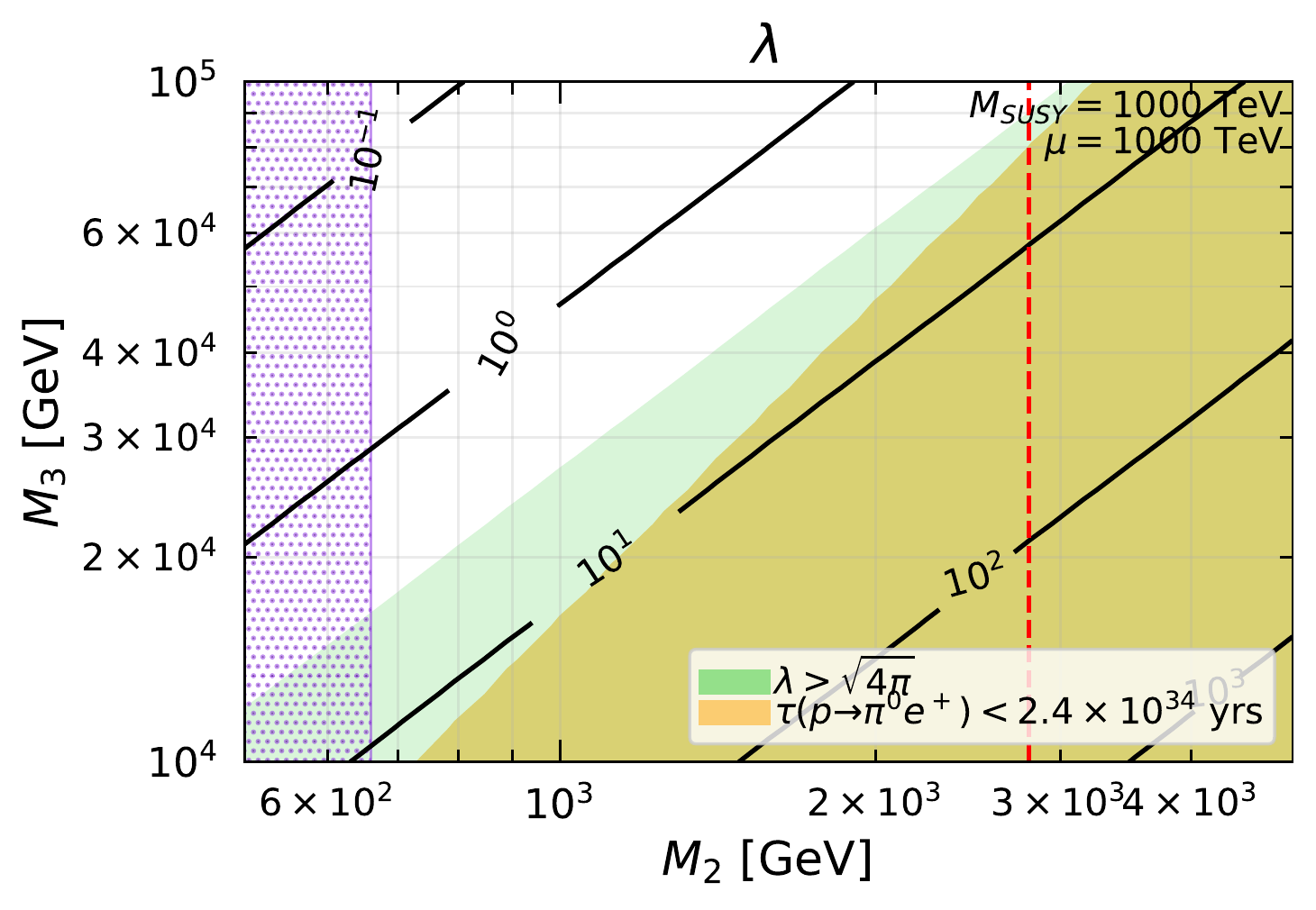}}
  \vspace{15pt}
  \subcaptionbox{\label{fig:alpha5}
  $\alpha_5^{-1} (Q_G)$
  }
  {\includegraphics[width=0.495\textwidth]{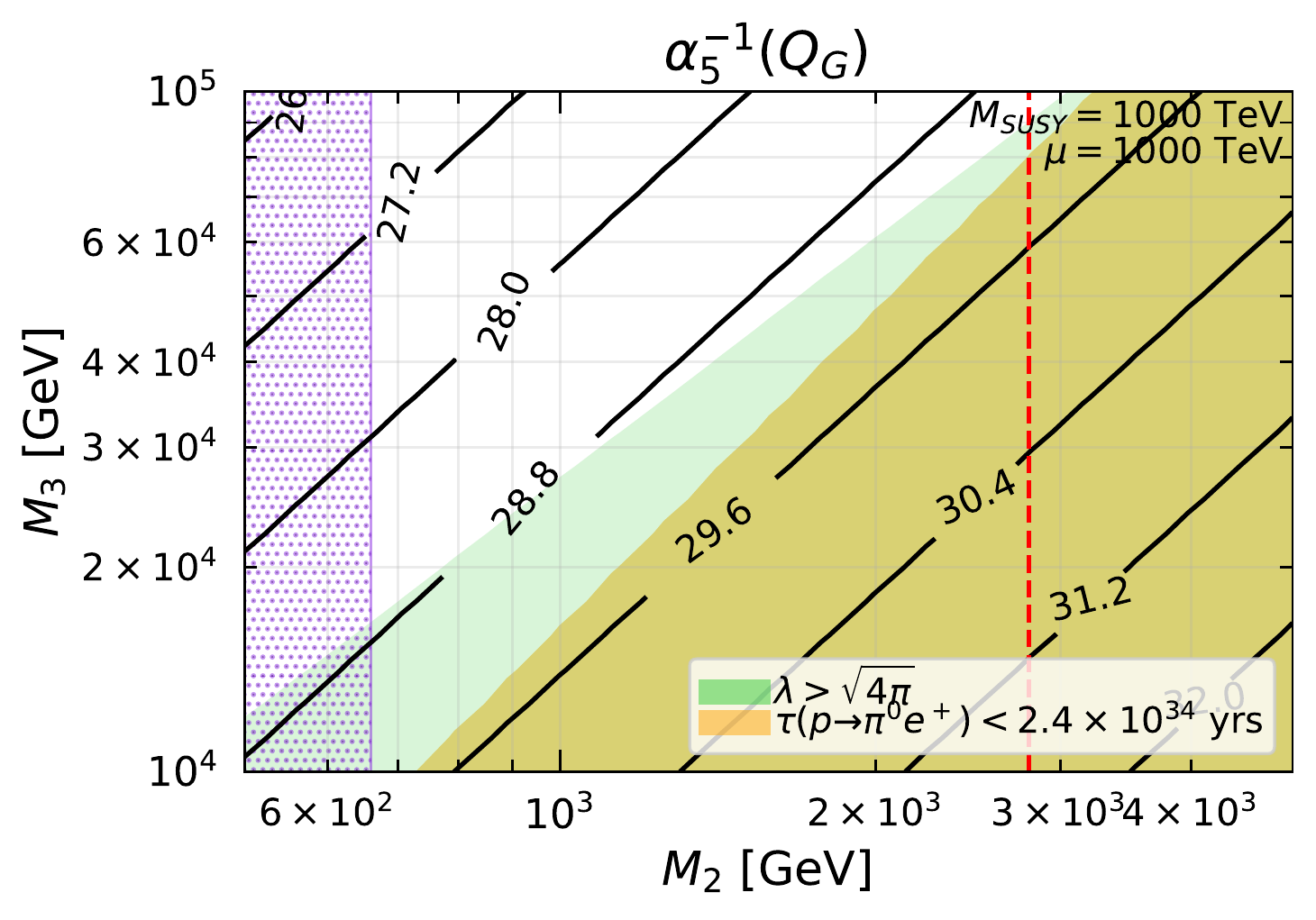}}
  \subcaptionbox{\label{fig:MH}
  $M_{H_C}$
  }
  {\includegraphics[width=0.495\textwidth]{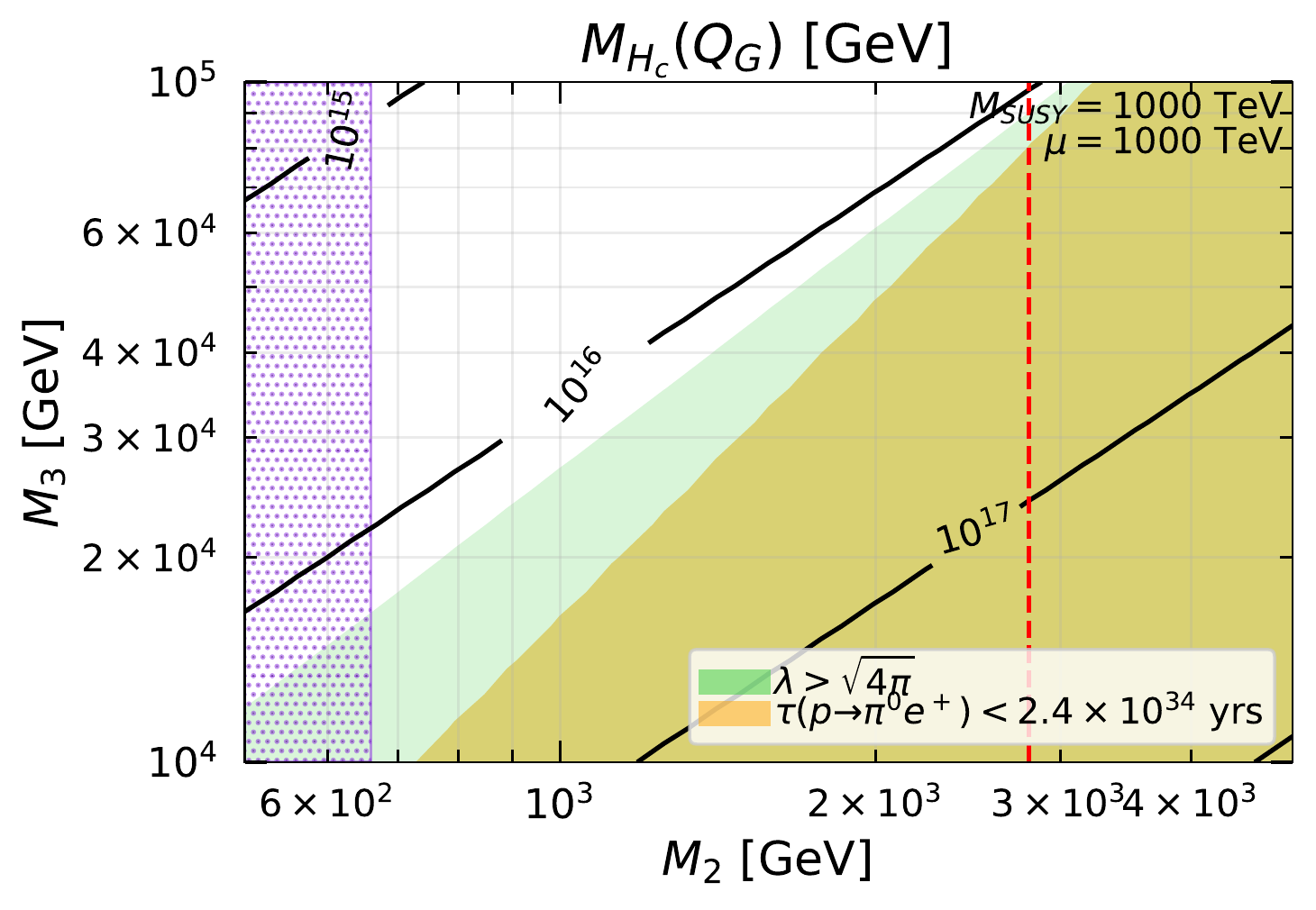}}
  \subcaptionbox{\label{fig:MX}
  $M_{X}$
  }
  {\includegraphics[width=0.495\textwidth]{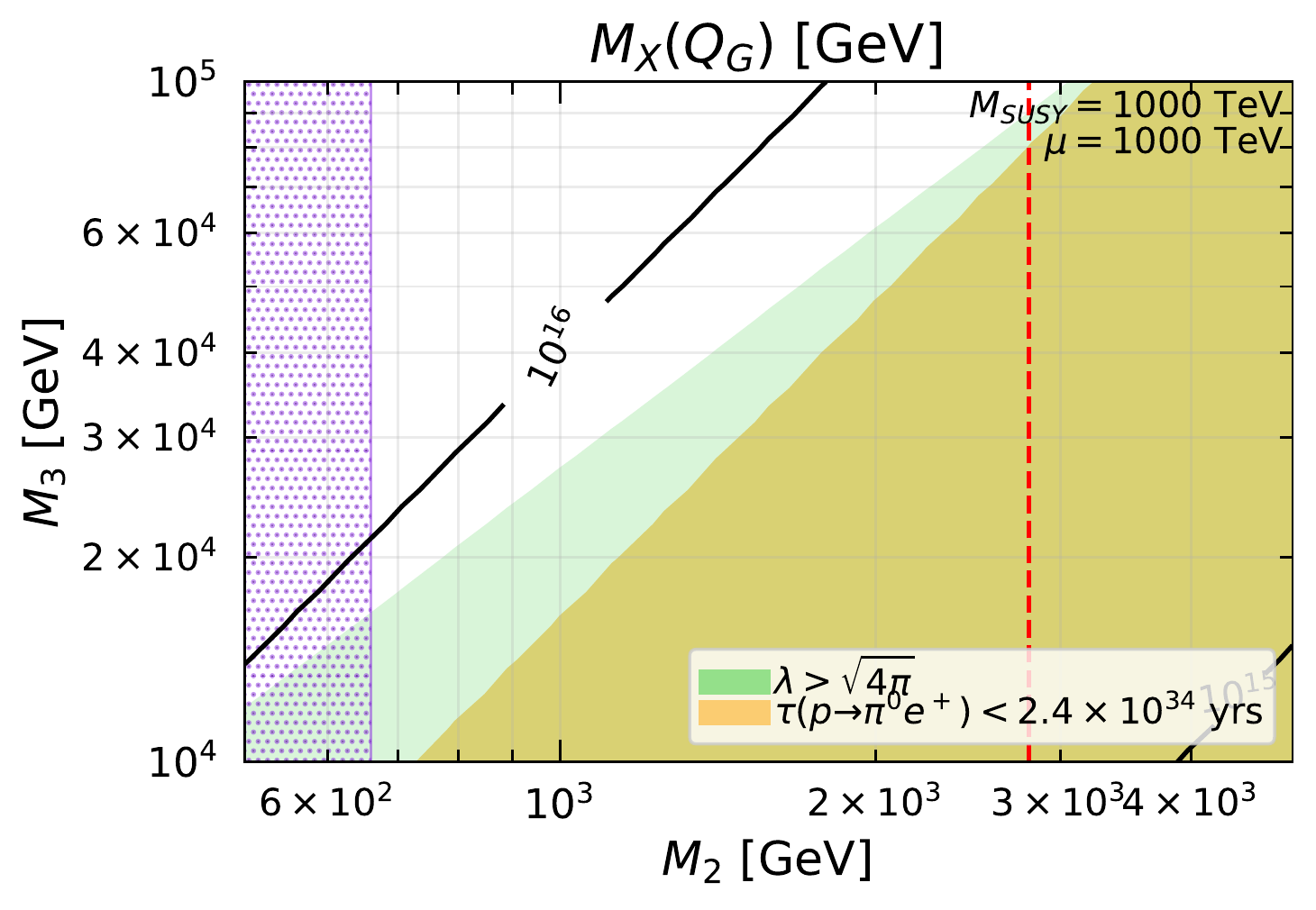}}
  \caption{
    Contour plots of (a) $\lambda$, (b) $\alpha_5^{-1} (Q_G)$, (c) $M_{H_C}$, and (d) $M_X$ in the $M_2$--$M_3$ plane in the minimal NG Higgs SUSY SU(5) GUT model. The masses of the SUSY scalar particles and higgsino are set to be $1~\mathrm{PeV}$ and $\tan \beta$ to be $\tan \beta = 3$. The green shaded region represents $\lambda > \sqrt{4\pi}$, which is ruled out by the perturbativity condition. The purple dotted region is excluded by the ATLAS disappearing track search~\cite{ATLAS:2022rme}. The yellow shaded area is excluded by the current experimental limit on $\tau (p \to e^+ \pi^0)$ from Super-Kamiokande~\cite{PhysRevD.102.112011}. The red vertical dashed line corresponds to the wino mass with which the observed dark matter density is reproduced with the thermal relic abundance of the wino dark matter. 
  }
\label{fig:m23}
\end{figure}
%%%%%%%%%%%%%%%%%%%%%%%%%%%%%%%%%%%%%%%%%%%%%%%%%%%%%%%%%%%%%%%%

Fig.~\ref{fig:m23} shows the contour plots of (a) $\lambda$, (b) $\alpha_5^{-1} (Q_G)$, (c) $M_{H_C}$, and (d) $M_X$ in the $M_2$--$M_3$ plane, where we set the masses of sfermions, heavy Higgs bosons, and higgsino to be $1~\mathrm{PeV}$ and $\tan \beta = 3$. $M_2$ and $M_3$ are soft SUSY-breaking wino mass and gluino mass, respectively. As seen from Fig.~\ref{fig:lambda}, the value of $\lambda$ can be much larger than unity for a large wino mass and/or a small gluino mass; this behavior can be understood from Fig.~\ref{fig:newlambda} and Eq.~\eqref{eq:alpha2alpha3}. In the green shaded regions in Fig.~\ref{fig:m23}, the coupling $\lambda$ is non-perturbative ($\lambda > \sqrt{4\pi}$). On the other hand, as seen in Fig.~\ref{fig:alpha5}, the SU(5) gauge coupling is predicted to be perturbative over the parameter region shown in this figure. The purple dotted region is excluded by the ATLAS disappearing track search: $m_{\widetilde{W}} > 660~\mathrm{GeV}$~\cite{ATLAS:2022rme}. The red vertical dashed line corresponds to the wino mass ($m_{\widetilde{W}} \simeq 2.8~\mathrm{TeV}$~\cite{Hisano:2006nn, Beneke:2020vff}) with which the observed dark matter density is reproduced with the thermal relic abundance of the wino dark matter. If $m_{\widetilde{W}} > 2.8~\mathrm{TeV}$, the wino relic is overabundant and we need some non-trivial cosmological history to dilute it. 

%%%%%%%%%%%%%%%%%%%%%%%%%%%%%%%%%%%%%%%%%%%%%%%%%%%%%%%%%%%%%%%%
\begin{figure}
  \centering
  \subcaptionbox{\label{fig:decay5_100}
  $\tau (p \to K^+ \bar{\nu})$, $M_{\mathrm{SUSY}} = 100~\mathrm{TeV}$
  }
  {\includegraphics[width=0.495\textwidth]{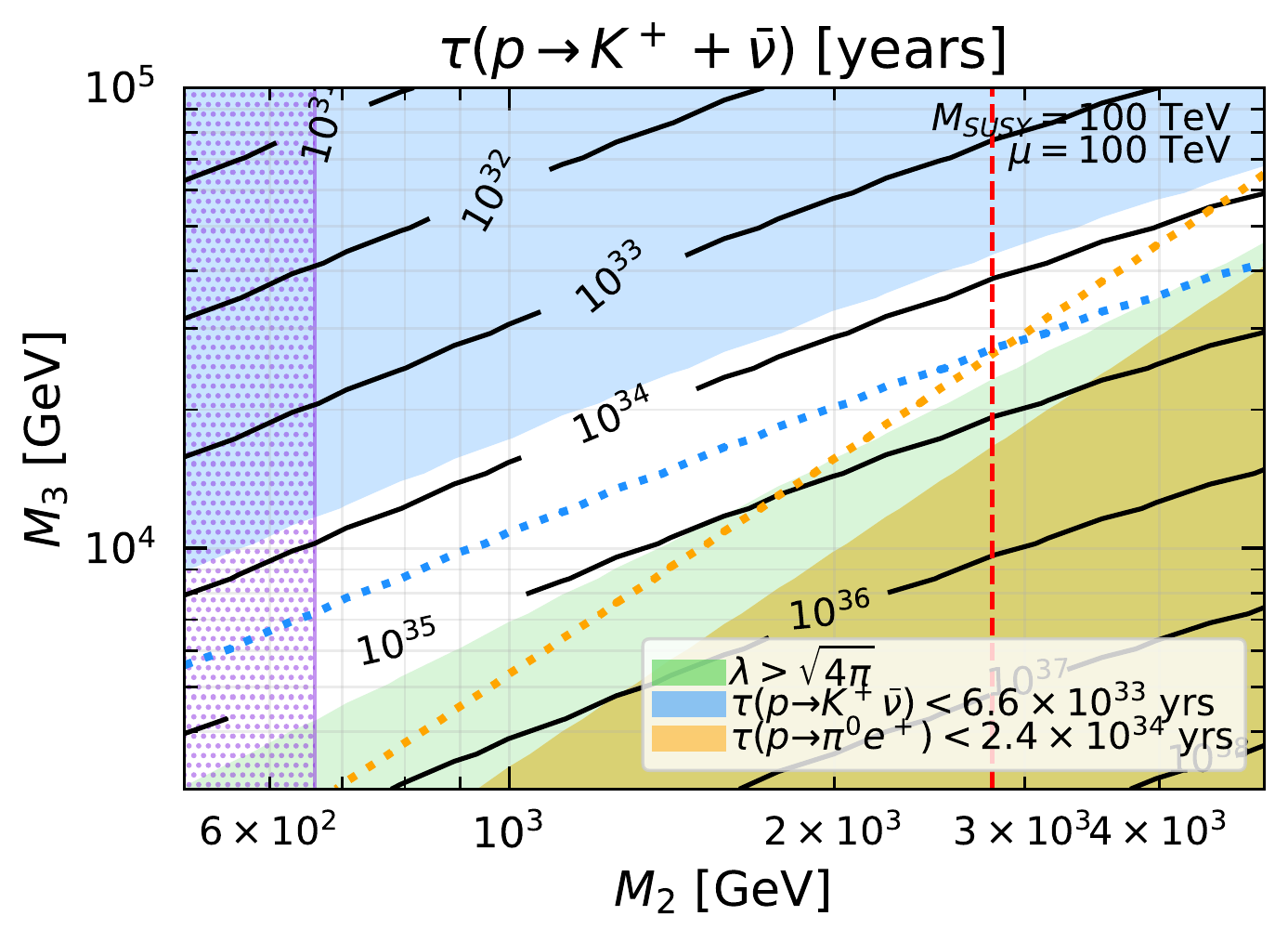}}
  \vspace{15pt}
  \subcaptionbox{\label{fig:decay6_100}
  $\tau (p \to e^+ \pi^0)$, $M_{\mathrm{SUSY}} = 100~\mathrm{TeV}$
  }
  {\includegraphics[width=0.495\textwidth]{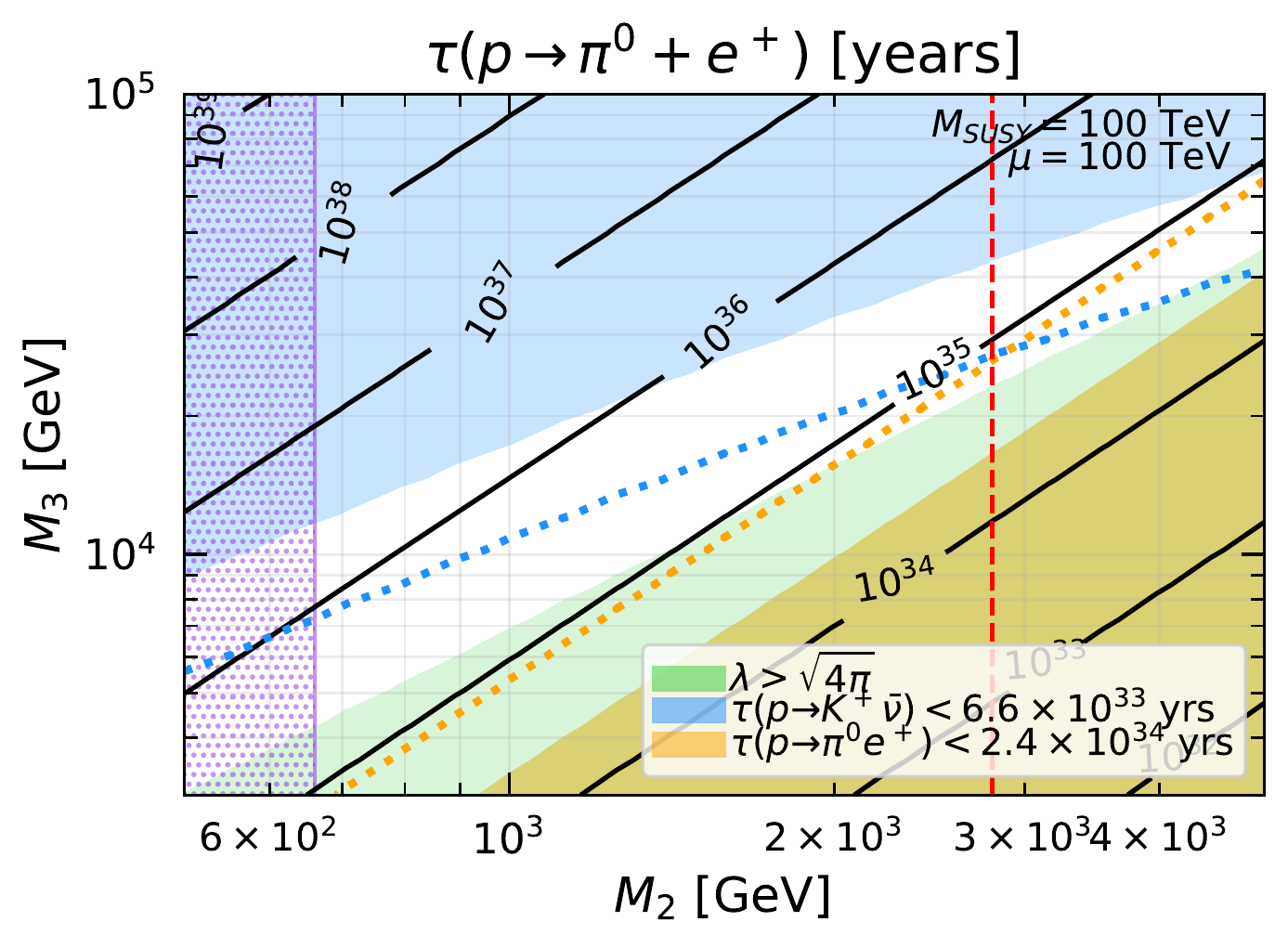}}
  \subcaptionbox{\label{fig:decay5_300}
  $\tau (p \to K^+ \bar{\nu})$, $M_{\mathrm{SUSY}} = 300~\mathrm{TeV}$
  }
  {\includegraphics[width=0.495\textwidth]{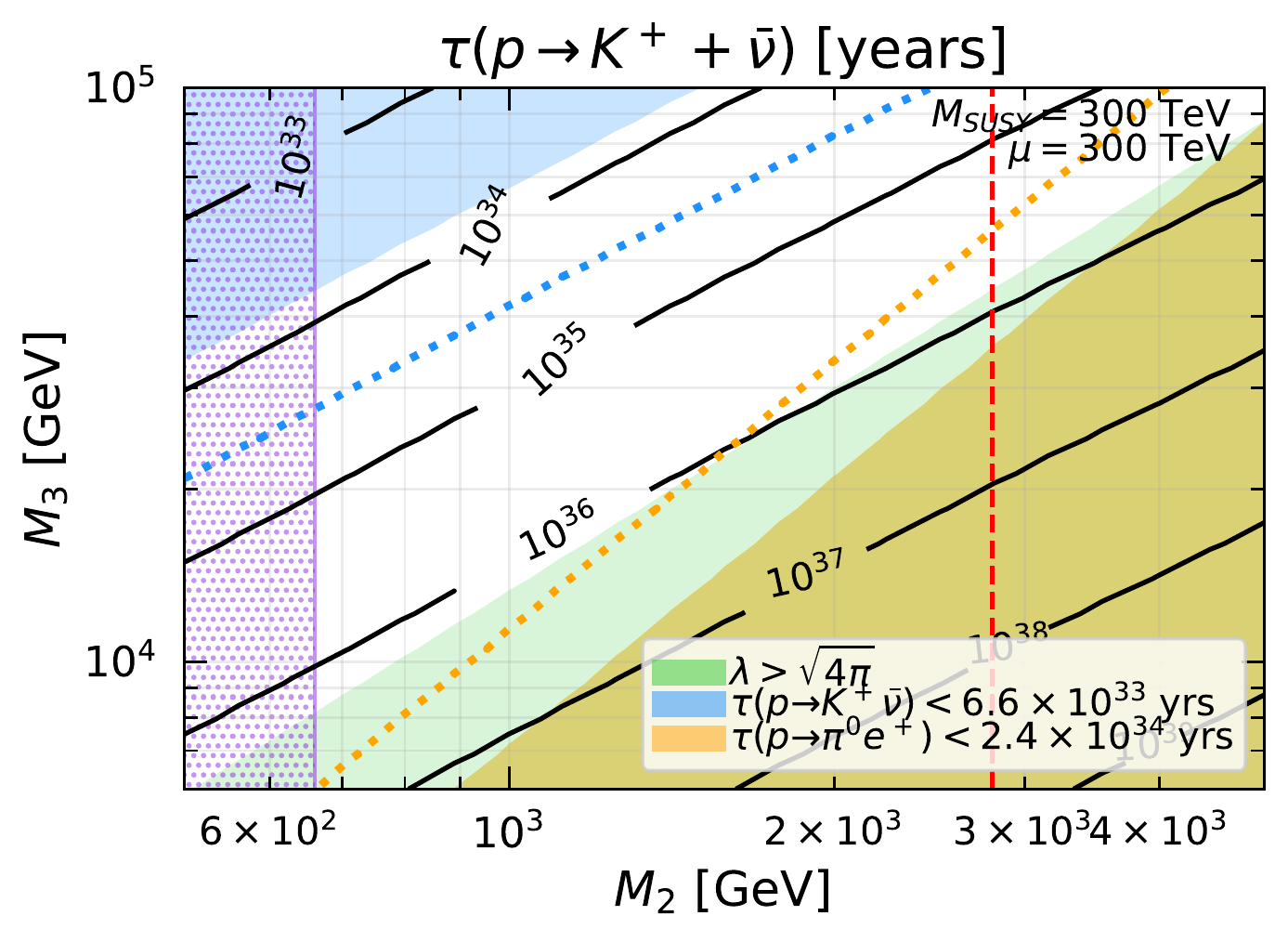}}
  \vspace{15pt}
  \subcaptionbox{\label{fig:decay6_300}
  $\tau (p \to e^+ \pi^0)$, $M_{\mathrm{SUSY}} = 300~\mathrm{TeV}$
  }
  {\includegraphics[width=0.495\textwidth]{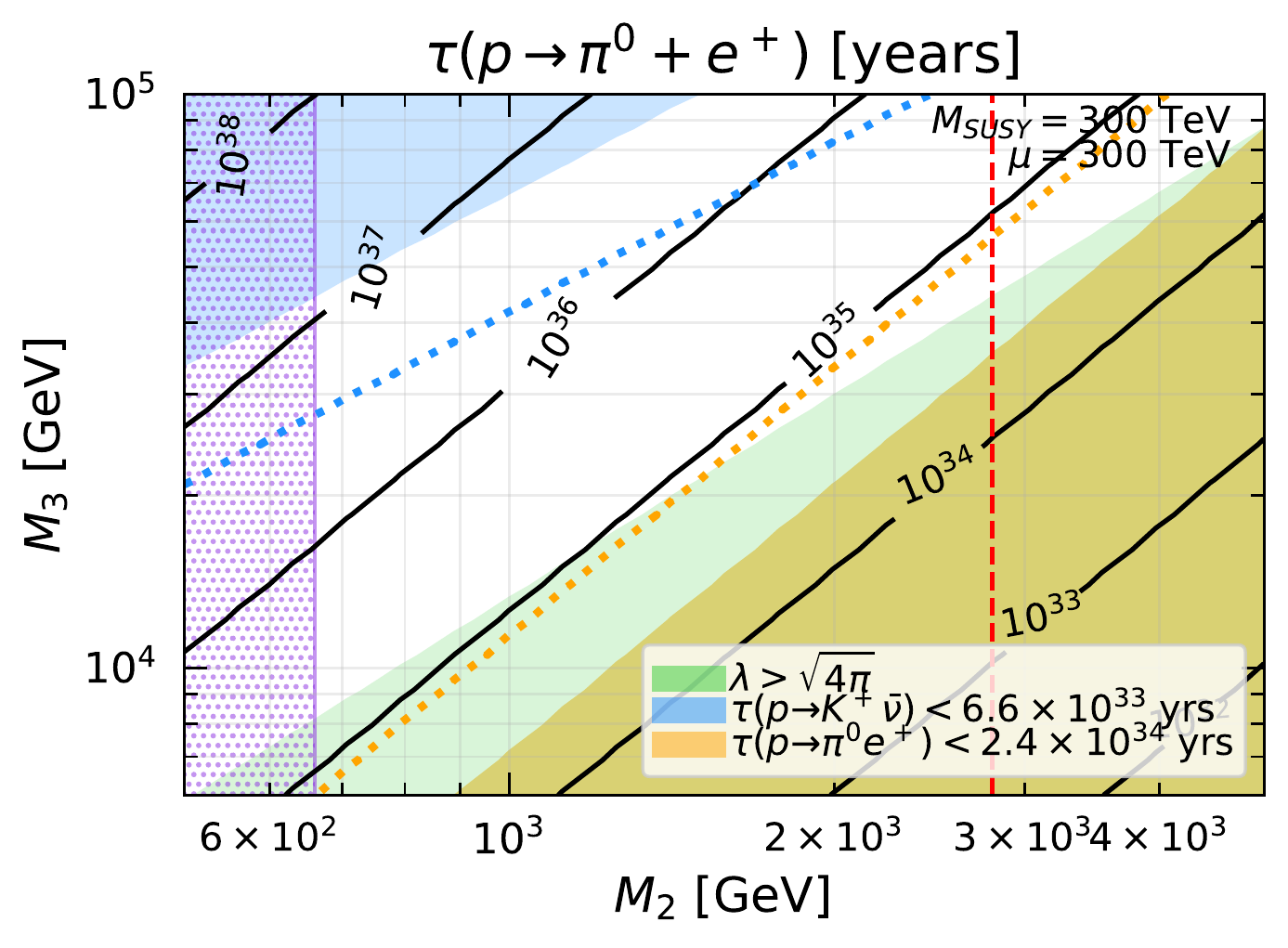}}
  \subcaptionbox{\label{fig:decay5_1000}
  $\tau (p \to K^+ \bar{\nu})$, $M_{\mathrm{SUSY}} = 1~\mathrm{PeV}$
  }
  {\includegraphics[width=0.495\textwidth]{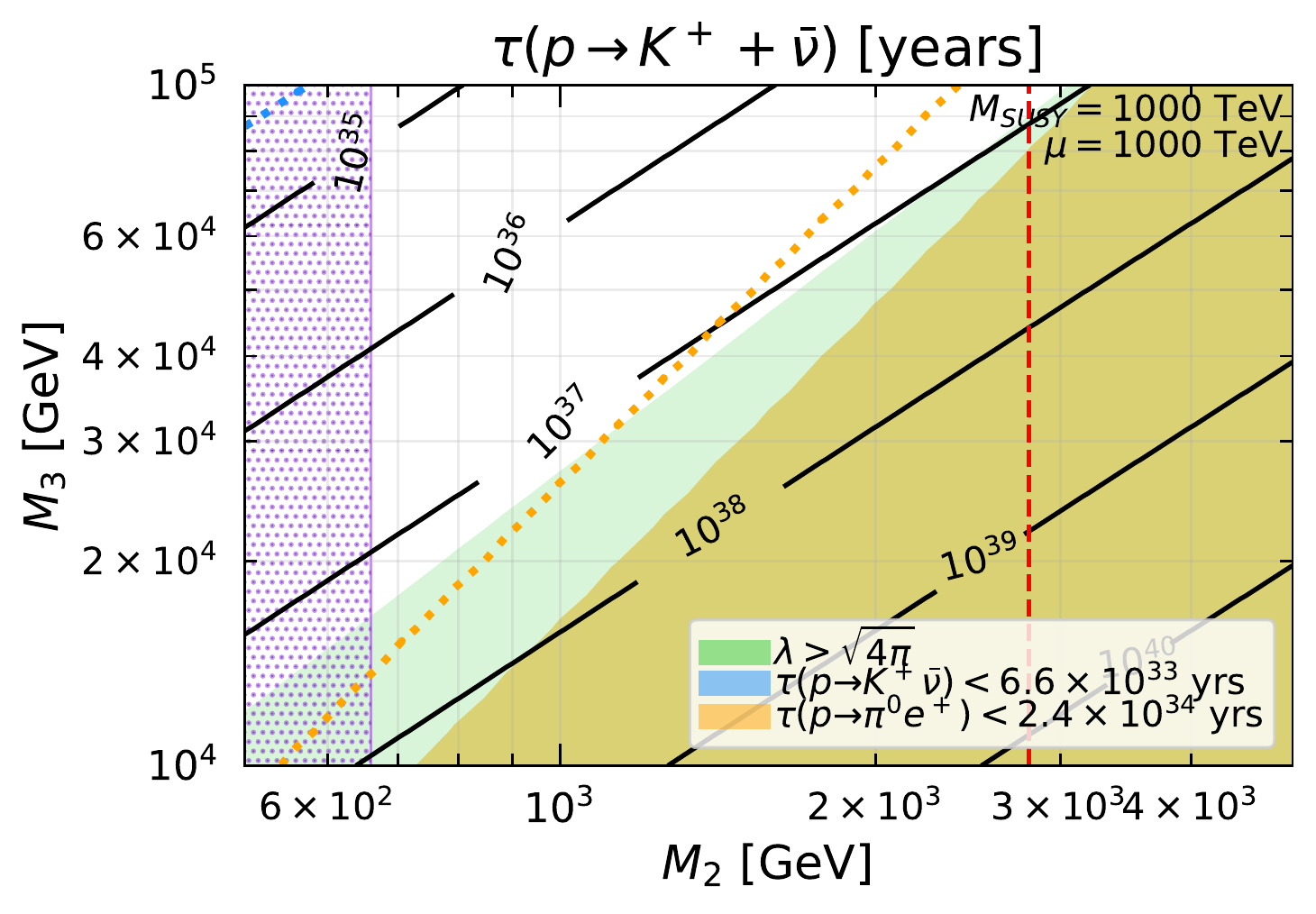}}
  \subcaptionbox{\label{fig:decay6_1000}
  $\tau (p \to e^+ \pi^0)$, $M_{\mathrm{SUSY}} = 1~\mathrm{PeV}$
  }
  {\includegraphics[width=0.495\textwidth]{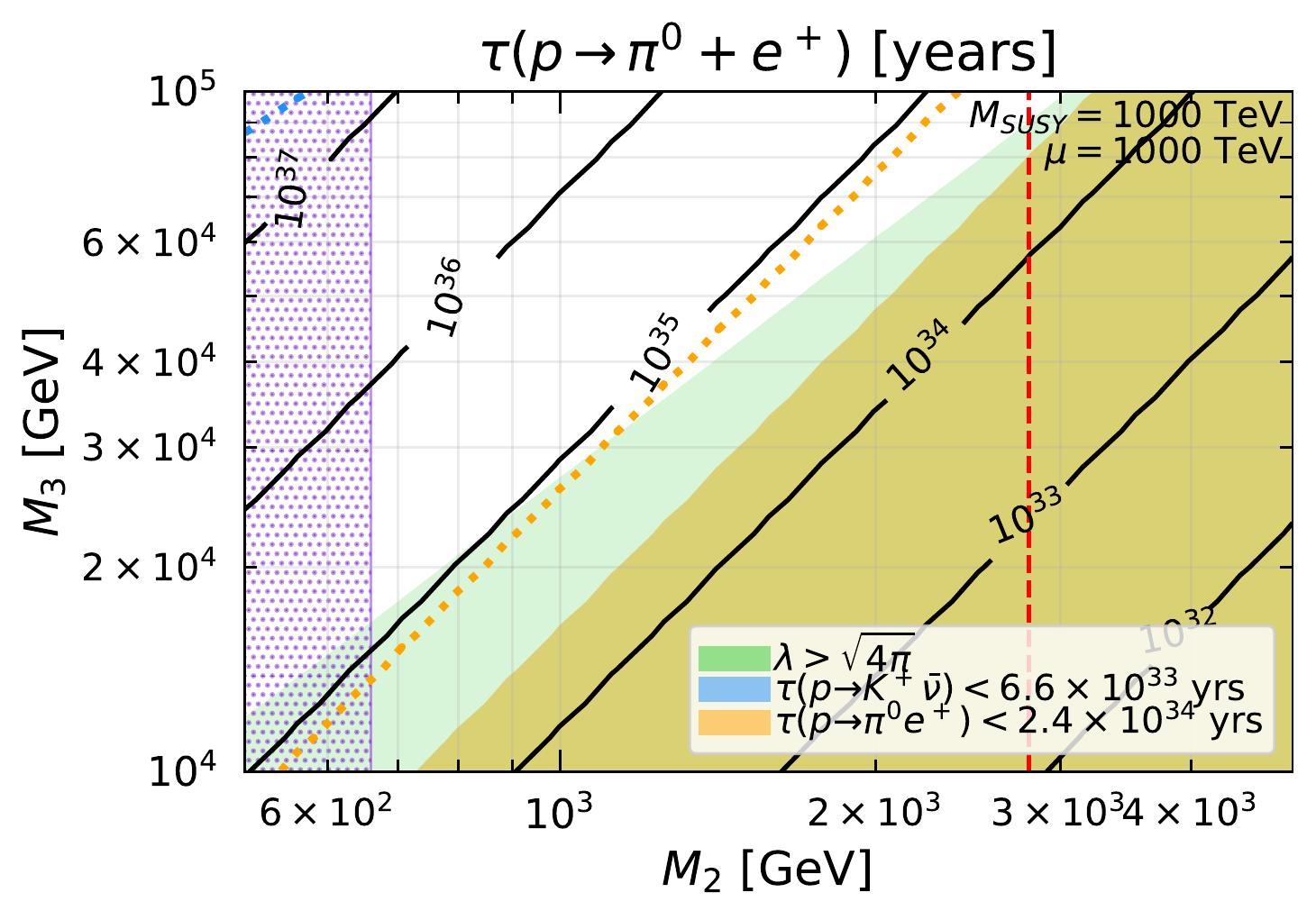}}
  \caption{
    Contour plots of proton lifetime for decay modes $\tau (p \to K^+ \bar{\nu})$ and $\tau (p \to e^+ \pi^0)$ in the $M_2$--$M_3$ plane in the minimal NG Higgs SUSY SU(5) GUT model. 
    The masses of the SUSY scalar particles and higgsino, $M_{\mathrm{SUSY}}$, are set to be 100 TeV, 300 TeV, and 1 PeV for (a)(b), (c)(d), and (e)(f), respectively, and $\tan\beta$ to be $\tan\beta=3$. The phases $\varphi_2$ and $\varphi_3$ are set to be $0$. The blue shaded area is excluded by the current experimental limit on $\tau (p \to K^+ \bar{\nu})$ from Super-Kamiokande~\cite{Super-Kamiokande:2014otb, Takhistov:2016eqm} while the yellow shaded area is excluded by the experimental limit on $\tau (p \to e^+ \pi^0)$~\cite{PhysRevD.102.112011}. The dotted lines show the sensitivities of Hyper-Kamiokande~\cite{Abe:2018uyc}. 
  }
\label{fig:ProtonDecay}
\end{figure}
%%%%%%%%%%%%%%%%%%%%%%%%%%%%%%%%%%%%%%%%%%%%%%%%%%%%%%%%%%%%%%%%

To satisfy the perturbativity condition for the $\lambda$ coupling, we need a small wino mass and a relatively large gluino mass, where the color-triplet Higgs mass is predicted to be $\lesssim 10^{16}$~GeV. With this size of $M_{H_C}$, the proton-lifetime limit can be relevant even in the high-scale SUSY scenario. In Fig.~\ref{fig:ProtonDecay}, we show contour plots of the proton lifetime of the dominant decay modes $\tau (p \to K^+ \bar{\nu})$ and $\tau (p \to e^+ \pi^0)$ in the $M_2$--$M_3$ plane with $M_{\mathrm{SUSY}} = 100~\mathrm{TeV}$, 300~TeV, and 1~PeV, the common masses of sfermions, heavy Higgs bosons, and higgsino. The phase factors $\varphi_2$ and $\varphi_3$ of the Yukawa couplings, which are defined in Sec.~\ref{sec:dim5}, are set to be $0$.  The blue and yellow shaded areas are ruled out by the current limits, $\tau (p \to K^+ \bar{\nu})>6.6\times 10^{33}~\mathrm{years}$~\cite{Super-Kamiokande:2014otb, Takhistov:2016eqm} and $\tau (p \to e^+ \pi^0)>2.4\times 10^{34}~\mathrm{years}$~\cite{PhysRevD.102.112011}, respectively, from Super-Kamiokande experiments. The expected 90\% CL limits from the 10-year run of Hyper-Kamiokande~\cite{Abe:2018uyc} are shown as
the blue and yellow dotted lines, which are $\tau (p \to K^+ \bar{\nu})>3.2\times 10^{34}~\mathrm{years}$ and $\tau (p \to e^+ \pi^0)>7.8\times 10^{34}~\mathrm{years}$, respectively.\footnote{Let us comment on the sensitivities of other future proton decay experiments~\cite{Dev:2022jbf}. The sensitivity of the JUNO experiment to $p \to K^+ \bar{\nu}$ is estimated to be $\tau (p \to K^+ \bar{\nu}) = 8.34 \times 10^{33}~\mathrm{years}$ for 10-year running~\cite{JUNO:2021vlw}, while that of DUNE is $\tau (p \to K^+ \bar{\nu}) = 1.3 \times 10^{34}~\mathrm{years}$~\cite{DUNE:2020fgq}. A relatively new proposal called THEIA~\cite{Theia:2019non, Theia:2022uyh} envisions a sensitivity of $\tau (p \to K^+ \bar{\nu})>3.8\times 10^{34}~\mathrm{years}$ with 800~kton-yrs data. For $p \to e^+ \pi^0$, the Hyper-Kamiokande experiment will offer much better sensitivity than those of the other experiments~\cite{Dev:2022jbf, Bhattiprolu:2022xhm}.
} As in Fig.~\ref{fig:m23}, the purple dotted region is excluded by the ATLAS disappearing track search and the red vertical dashed line corresponds to the mass of the thermal wino dark matter. 
We find that the perturbativity condition on $\lambda$ and bound from $\tau (p \to e^+ \pi^0)$ give upper and lower limits on the wino and gluino masses, respectively. On the other hand, the bound from $\tau (p \to K^+ \bar{\nu})$ gives restrictions on small wino mass and large gluino mass. Consequently, they together give a belt of allowed parameter space in Fig.~\ref{fig:ProtonDecay}. The boundary of the belt depends on $M_{\rm SUSY}$; when $M_{\mathrm{SUSY}}$ gets larger, the limit from $\tau (p \to K^+ \bar{\nu})$ becomes weaker, while the limits from $\tau (p \to e^+ \pi^0)$  and the perturbativity condition on $\lambda$ get stronger. 
For $M_{\rm SUSY} = 100~{\mathrm{TeV}}$, a large part of the region motivated by the wino dark matter scenario, $M_2 \leq 2.8~\mathrm{TeV}$, has already been excluded by these limits, and most of the allowed region can be probed by the Hyper-Kamiokande experiment. In addition, the perturbativity bound on $\lambda$ gives a lower limit on the gluino mass, $M_3 \gtrsim 4~\mathrm{TeV}$, which is stronger than the current LHC limit~\cite{CMS:2019zmd, ATLAS:2020syg}. For $M_{\rm SUSY} = 1~{\mathrm{PeV}}$, on the other hand, a sizable region of the parameter space is still allowed, mainly because the $\tau (p \to K^+ \bar{\nu})$ bound is fairly weak in this case. A part of this allowed region can be probed at the Hyper-Kamiokande through the $\tau (p \to e^+ \pi^0)$ channel. Notice that the gluino mass $M_3$ needs to be smaller than the SUSY-breaking scale by at least an order of magnitude in the present scenario since it is supposed to be generated by radiative corrections. This theoretical requirement imposes a limit $M_3 \lesssim 100~\mathrm{TeV}$ for $M_{\rm SUSY} = 1~{\mathrm{PeV}}$.\footnote{For instance, in the anomaly mediation, $|M_3| \simeq 3 \alpha_3 M_{\rm SUSY}/(4\pi)$, and thus $|M_3| \sim 30~\mathrm{TeV}$ for $M_{\rm SUSY} = 1~{\mathrm{PeV}}$.} This entire region can be probed if (far) future experiments can have sensitivities to the lifetime of $\mathcal{O}(10^{36})$~years in both the $p \to K^+ \bar{\nu}$ and $p \to e^+ \pi^0$ channels. 

Let us finally comment on the uncertainty in our calculation of proton lifetimes resulting from those of the SM parameters. It turns out that the uncertainty of $\alpha_3$ gives the largest influence on our results. It can induce a factor $2$ difference to $\lambda$ and $\mathcal{O}(1)$ difference to proton decay widths. The uncertainties resulting from the hadron matrix elements are ${\cal O}(0.1)$ (cf. App.~\ref{sec:protondecay}), and the effects of other parameters' uncertainties are found to be less than $\mathcal{O}(10^{-2})$, which are subdominant.
For example, for the point at $M_2=1$~TeV and $M_3=50$~TeV with $M_{\mathrm{SUSY}}=1$~PeV in Fig.~\ref{fig:decay5_1000}, 
$\alpha_3=0.1179\pm 0.0009$ 
gives 
$\lambda=0.86^{+0.50}_{-0.31}$,
$\tau (p \to K^+ \bar{\nu})=2.2^{+2.2}_{-1.1}\times 10^{36}$~years, 
and 
$\tau (p \to e^+ \pi^0)=4.1^{+2.7}_{-1.7}\times 10^{35}$~years.
In addition, the complex phases $\varphi_2$ and $\varphi_3$ in the GUT Yukawa coupling, which are defined in Eq.~\eqref{eq:dim5gutmatchijkl} in Appendix~\ref{sec:dim5}, 
can also change the lifetime of proton decay mode $p \to K^+ \bar{\nu}$. For the model point $(M_2, M_3,M_{\mathrm{SUSY}})=$(1~TeV, 50~TeV, 1~PeV) in Fig.~\ref{fig:decay5_1000}, the lifetime varies 
$\tau (p \to K^+ \bar{\nu})=(2.18$ -- $2.32)\times 10^{36}~\mathrm{years}$ for $\varphi_2, \varphi_3 = 0$ -- $2\pi$.
For the model point $(M_2, M_3,M_{\mathrm{SUSY}})=$(2~TeV, 20~TeV, 100~TeV) in Fig.~\ref{fig:decay5_100}, the lifetime varies 
$\tau (p \to K^+ \bar{\nu})=(3.1$ -- $9.2)\times 10^{34}~\mathrm{years}$ for $\varphi_2, \varphi_3 = 0$ -- $2\pi$. As we see, the uncertainty from the GUT-Yukawa phases reduces for larger $M_{\rm SUSY}$. This is because for large $M_{\rm SUSY} = \mu$, the higgsino exchange contribution dominates the wino contribution since $\mu \gg M_2$, and in this case the decay amplitude depends mainly on the overall phase factor $e^{i (\varphi_2 + \varphi_3)}$, making the decay rate almost independent of the GUT phases.

%%%%%%%%%%%%%%%%%%%%%%%%%%%%%%%%%%%%%%%%%%%%%%%%%%%%%%%%
\section{Conclusions and discussion}
\label{sec:summary}
%%%%%%%%%%%%%%%%%%%%%%%%%%%%%%%%%%%%%%%%%%%%%%%%%%%%%%%%

In this paper, we have studied the minimal NG Higgs SUSY SU(5) GUT model, which can naturally lead to light Higgs doublets and solve the doublet-triplet splitting problem. The model is more restrictive than the usual SU(5) models and has sharper predictions. We determine all the GUT parameters by using the matching conditions for the SM gauge coupling constants at the GUT scale. We then find that the perturbativity condition on the coupling $\lambda$ sets constraints $\alpha_2^{-1} (Q_G) \gtrsim \alpha_3^{-1} (Q_G) $ and $M_{H_C} \lesssim M_X$. The former constraint restricts the allowed pattern of the low-energy SUSY mass spectrum through RGEs; we see that this gives upper limits to the wino mass $m_{\widetilde{W}}$, higgsino mass $m_{\widetilde{H}}$, and heavy Higgs doublet mass $m_A$, and a lower limit to the gluino mass $m_{\widetilde{g}}$.

To see the implications of these constraints for the low-energy SUSY spectrum, we have studied two concrete SUSY scenarios: the CMSSM and high-scale SUSY. For the CMSSM, we have found that the perturbativity condition on $\lambda$ leads to a limit on the color triplet-Higgs mass, $M_{H_C} \lesssim 10^{16}$~GeV, and this is too low to evade the experimental limit on $p \to K^+ \bar{\nu}$, as the SUSY scale is predicted to be $\mathcal{O}(10)$~TeV in this scenario. This problem can be evaded in the high-scale SUSY scenario. We have seen that the present bound on $p \to K^+ \bar{\nu}$ gives a lower (upper) limit on the wino (gluino) mass. On the other hand, the perturbativity condition on $\lambda$, as well as the bound on $p \to e^+ \pi^0$, gives an upper (lower) limit on $M_2$ ($M_3$). As a result, we found an allowed band in the $M_2$--$M_3$ plane for a given value of $M_{\mathrm{SUSY}}$. For $M_{\mathrm{SUSY}} = 100~\mathrm{TeV}$, most of the remaining parameter space can be explored at the Hyper-Kamiokande. For a higher SUSY-breaking scale, the constraints from proton decay searches are relaxed and a considerable part of the parameter space is beyond the reach of the Hyper-Kamiokande experiments. To cover the whole allowed region in the case of $M_{\mathrm{SUSY}} = 1~\mathrm{PeV}$, for instance, we need sensitivities to the proton lifetimes of $\mathcal{O}(10^{36})~\mathrm{years}$ for both $p \to K^+ \bar{\nu}$ and $p \to e^+ \pi^0$. 

The high-scale SUSY scenario in the minimal NG Higgs SUSY SU(5) GUT can also be tested at future colliders. As we have seen, a relatively light wino mass is favored in this setup and thus can be a good target for future collider experiments. The HL-LHC is expected to probe the wino mass up to $\simeq 850$~GeV~\cite{ATLAS:2018jjf}. This reach can be extended significantly at FCC-hh, which is expected to exceed the thermal wino dark matter mass $m_{\widetilde{W}} \simeq 2.8$~TeV~\cite{Saito:2019rtg}. The FCC-hh can also probe gluinos with a mass $\lesssim 13$~TeV~\cite{Golling:2016gvc}. The discovery (or exclusion) of these gauginos offers another way of checking the validity of our model; for instance, if their masses are measured,\footnote{See Ref~\cite{Asai:2019wst} for the prospects of gaugino mass measurements at a 100~TeV $pp$ collider. } we can predict the proton-decay lifetimes as functions of $M_{\mathrm{SUSY}}$, which can be tested in future proton-decay experiments. 

Although we have focused on the minimal NG Higgs SUSY SU(5) GUT in this paper, the same analysis can also be performed in other non-minimal NG Higgs GUT scenarios, such as the SU(6) GUT model with an $\mathrm{SU}(6) \otimes \mathrm{SU}(6)$ global symmetry~\cite{Berezhiani:1989bd, Barbieri:1992yy, Barbieri:1993wz, Barbieri:1994kw, Berezhiani:1995sb, Berezhiani:1995dt}. We expect that the presence of a global symmetry again gives non-trivial relations among the GUT parameters and thus makes the model highly predictive. We note in passing that NG Higgs GUT models may be related to extra-dimensional GUT models (see, \textit{e.g.}, Refs.~\cite{Hall:2001zb, Burdman:2002se, Haba:2004qf, Lim:2007jv, Angelescu:2021nbp} for such examples), as suggested by the AdS/CFT correspondence~\cite{Maldacena:1997re}, and therefore we may indirectly explore these extra-dimensional GUT models by probing NG Higgs GUT models in future experiments.

%%%%%%%%%%%%%%%%%%%%%%%%%%%%%%%%%%%%
\section*{Acknowledgments}
%%%%%%%%%%%%%%%%%%%%%%%%%%%%%%%%%%%%

This work is supported in part by the Grant-in-Aid for Innovative Areas (No.19H05810 [KH], No.19H05802 [KH], No.18H05542 [NN]), Scientific Research B (No.20H01897 [KH and NN]), Young Scientists (No.21K13916 [NN]), and JSPS Fellowship (No.22J20755 [SH]). SH would like to thank Maura E. Ramirez-Quezada for the useful advice and comments.

%%%%%%%%%%%%%%%%%%%%%%%%%%%%%%%%%%%%%%%%%%%%%%
\section*{Appendix}
\appendix
%%%%%%%%%%%%%%%%%%%%%%%%%%%%%%%%%%%%%%%%%%%%%

%%%%%%%%%%%%%%%%%%%%%%%%%%%%%%%%%%%%%%%%%%%%%%
\section{Renormalization group analysis}
\label{sec:RG}
\renewcommand{\theequation}{A.\arabic{equation}}
\setcounter{equation}{0}
%%%%%%%%%%%%%%%%%%%%%%%%%%%%%%%%%%%%%%%%%%%%%%

%%%%%%%%%%%%%%%%%%%%%%%%%%%%%%%%%
\subsection{Input parameters}
\label{sec:input}
%%%%%%%%%%%%%%%%%%%%%%%%%%%%%%%%%

%%%%%%%%%%%%%%%%%%%%%%%%%%%%%%%%%%%%%%%%%%%%%%%%%%%%%%%%%%%%%%%%%%%%%%%%%%%%%%%5
\begin{table}[t]
  \centering
  \caption{Input parameters~\cite{ParticleDataGroup:2020ssz}.}
%  \vspace{3mm}
  \begin{tabular}{ll}
    \hline\hline
    Fermi constant  $G_F$ & $1.1663788 (6) \times 10^{-5}~\mathrm{GeV}^{-2}$ \\ 
    Strong coupling constant  $\alpha_s$ & $0.1179(9)$ \\ 
    $Z$ boson mass $m_Z$ & $91.1876 (21)~\mathrm{GeV}$ \\ 
    $W$ boson mass $m_W$ & $80.377(12)~\mathrm{GeV}$ \\ 
    Higgs boson mass $m_h$ & $125.25(17)~\mathrm{GeV}$ \\ 
    Top quark mass $m_t$ & $172.69(30)~\mathrm{GeV}$ \\ 
    Bottom quark mass $m_b$ & $4.18^{+0.03}_{-0.02}~\mathrm{GeV}$ \\ 
    Tau lepton mass $m_{\tau}$ & $1.77686(12)~\mathrm{GeV}$ \\ 
    Charm quark mass $m_{c}$ & $1.27(2)~\mathrm{GeV}$ \\ 
    Strange quark mass $m_{s}$ & $93.4^{+8.6}_{-3.4}~\mathrm{MeV}$ \\ 
    Muon mass $m_{\mu}$ & $105.658~\mathrm{MeV}$ \\ 
    Up quark mass $m_{u}$ & $2.16^{+0.49}_{-0.26}~\mathrm{MeV}$ \\ 
    Down quark mass $m_{d}$ & $4.67^{+0.48}_{-0.17}~\mathrm{MeV}$ \\ 
    Electron mass $m_{e}$ & $0.511~\mathrm{MeV}$ \\ 
    \hline\hline
  \end{tabular}
  \label{tab:input}
\end{table}
%%%%%%%%%%%%%%%%%%%%%%%%%%%%%%%%%%%%%%%%%%%%%%%%%%%%%%%%%%%%%%%%%%%%%%%%%%%%%%%%

We summarize the values of the input parameters in Table~\ref{tab:input}, which we take from Ref.~\cite{ParticleDataGroup:2020ssz}. The bottom and charm quark masses are the $\overline{\mathrm{MS}}$ masses renormalized at $m_b$ and $m_c$, respectively, and the up, down, and strange quark masses are the $\overline{\mathrm{MS}}$ masses at $2~\mathrm{GeV}$. The other masses are the pole masses. $\alpha_s$ is in the $\overline{\mathrm{MS}}$ scheme renormalized at $m_Z$.

By using the input parameters in Table~\ref{tab:input}, we obtain the $\overline{\mathrm{MS}}$ couplings at the scale of $m_t$. For the gauge couplings $g'$, $g_2$, and $g_3$ and the top Yukawa coupling $y_t$, we use the results given in Ref.~\cite{Buttazzo:2013uya}.

%%%%%%%%%%%%%%%%%%%%%%%%%%%%%%%%%%%%%%%%%%%%%%%%%%%%
%\subsubsection{QCD Corrections}
%\label{sec:qcd}
%%%%%%%%%%%%%%%%%%%%%%%%%%%%%%%%%%%%%%%%%%%%%%%%%%%%

We extract the quark Yukawa couplings except for $y_t$ from the $\overline{\mathrm{MS}}$ masses at the scale of $m_t$, which are obtained with the QCD RGEs at the two-loop level. For quark masses, we use 
\begin{equation}
  \frac{dm(Q)}{d \ln Q} =\biggl[ \gamma_{m(1)}\frac{\alpha_3}{4\pi} +\gamma_{m(2)}\frac{\alpha_3^2}{(4\pi)^2} \biggr] m(Q)~,
\end{equation}
where
\begin{equation}
  \gamma_{m(1)}=-6C_F~,\quad \gamma_{m(2)}= -C_F \biggl( 3C_F+\frac{97}{3}N_c -\frac{10}{3}N_f \biggr)~, \quad C_F \equiv \frac{N_c^2-1}{2N_c}~,
\end{equation}
with the number of colors $N_c=3$, the number of effective quark flavors $N_f$, and the quadractic Casimir invariant $C_F$. The RGE for the strong coupling constant is
\begin{equation}
    \frac{d\alpha_3}{d \ln Q}= (2b_1)\frac{\alpha_3^2}{4\pi}+ (2b_2)\frac{\alpha_3^3}{(4\pi)^2}~,
\end{equation}
with 
\begin{equation}
    b_1=-\frac{11N_c-2N_f}{3}~,\quad b_2 =-\frac{34}{3}N_c^2 +\frac{10}{3}N_cN_f +2C_FN_f~.\
\end{equation}

On the other hand, we determine the lepton Yukawa couplings from their pole masses using the tree-level relations.

%%%%%%%%%%%%%%%%%%%%%%%%%%%%%%%%%%%%%%%%%%%%%%%%%%%%
\subsection{Renormalization scheme transformation}
\label{sec:rensch}
%%%%%%%%%%%%%%%%%%%%%%%%%%%%%%%%%%%%%%%%%%%%%%%%%%%%

We convert the $\overline{\mathrm{MS}}$ couplings into the $\overline{\mathrm{DR}}$ couplings at the scale of $Q = m_t$. For the gauge coupling constants, we use the one-loop relations~\cite{Antoniadis:1982vr}
\begin{equation}
  \alpha_a (Q) |_{\overline{\mathrm{DR}}}  
  = \alpha_a (Q) |_{\overline{\mathrm{MS}}} 
  \biggl(1 + \frac{C(G_a)}{12\pi} \alpha_a (Q) |_{\overline{\mathrm{MS}}}  \biggr) ~, 
\end{equation}
where $C(G_a)$ is the quadratic Casimir invariant with $C(G_a) = 0, 2, 3$ for $a = 1,2,3$, respectively. On the other hand, we do not perform the scheme transformation for the Yukawa couplings since they appear in the RGEs of the gauge coupling constants only at the two-loop level.

%%%%%%%%%%%%%%%%%%
\subsection{RGEs}
\label{sec:RGEs}
%%%%%%%%%%%%%%%%%%

We use the two-loop RGEs to evolve the gauge coupling constants, which have the form  
\begin{equation}
   \frac{d g_a}{d \ln Q}=\frac{1}{16\pi^2}b_a 
 ^{(1)}g^3_a
 +\frac{g_a^3}{(16\pi^2)^2}\biggl[
 \sum_{b=1}^{3}b_{ab}^{(2)}g_b^2 -\sum_{k=u,d,e}c_{ak}~ {\rm
 Tr}(y^\dagger_k y_k^{}) 
 \biggr]~,
 \label{eq:rgegauge}
\end{equation}
where $y_k$ is the Yukawa matrix. Above the electroweak scale and below the gaugino mass scale, $Q_{\mathrm{gaugino}} = \sqrt{M_2 M_3}$, the coefficients are 
\begin{align}
  b_a^{(1)}&=
 \begin{pmatrix}
  41/10 \\ -19/6 \\ -7
 \end{pmatrix}
~, \quad 
  b^{(2)}_{ab} =
 \begin{pmatrix}
  199/50 & 27/10 & 44/5 \\
  9/10 & 35/6 & 12 \\
  11/10 & 9/2 & -26
 \end{pmatrix}
 ~,
 \quad 
  c_{ak}=
 \begin{pmatrix}
  17/10 & 1/2 & 3/2 \\
  3/2 & 3/2 & 1/2 \\
  2 & 2 & 0
\end{pmatrix}
~.
\end{align} 
Above the gaugino mass scale $Q_{\mathrm{gaugino}}$ and below the SUSY scale $Q_{\mathrm{SUSY}}$, we have additional contributions to the gauge coupling beta functions by gauginos: 
\begin{equation}
 \Delta b_a^{(1)}=
 \begin{pmatrix}
  0 \\ 4/3 \\ 2
\end{pmatrix}
 ~,
\qquad 
  \Delta b^{(2)}_{ab} =
 \begin{pmatrix}
  0 & 0 & 0 \\
  0 & 64/3 & 0 \\
  0 & 0 & 48
\end{pmatrix}
 ~.
\end{equation}
Above the SUSY scale $Q_{\mathrm{SUSY}}$, the coefficients in Eq.~\eqref{eq:rgegauge} are 
\begin{equation}
  b^{(1)}_a=
  \begin{pmatrix}
   33/5 \\ 1 \\ -3
  \end{pmatrix}
  ~,\quad b_{ab}^{(2)}=
  \begin{pmatrix}
   199/25 & 27/5 & 88/5 \\
   9/5 & 25 & 24 \\
   11/5 & 9 & 14 
  \end{pmatrix} 
  ~,\quad 
   c_{ak}=
  \begin{pmatrix}
   26/5 & 14/5 & 18/5 \\
   6 & 6 & 2 \\
   4 & 4 & 0
  \end{pmatrix}
  ~.
\end{equation}

For the Yukawa couplings, on the other hand, we use one-loop RGEs since they appear in the gauge-coupling RGEs only at the two-loop level. Below the SUSY scale $Q_{\mathrm{SUSY}}$, we have 
\begin{align}
  \frac{d y_u^{}}{d \ln Q}&=\frac{1}{16\pi^2}
 \biggl[
 \frac{3}{2}(y_u^{}y_u^\dagger-y_d^{}y_d^\dagger) +Y_2
  -\frac{17}{20}g^2_1-\frac{9}{4}g_2^2-8
  g_3^2
 \biggr]y_u^{} , \nonumber \\
  \frac{d y_d^{}}{d \ln Q}&=\frac{1}{16\pi^2}
 \biggl[
 \frac{3}{2}(y_d^{}y_d^\dagger-y_u^{}y_u^\dagger) +Y_2
  -\frac{1}{4}g^2_1-\frac{9}{4}g_2^2-8
  g_3^2
 \biggr]y_d^{} , \nonumber \\
  \frac{dy_e^{}}{d \ln Q}&=\frac{1}{16\pi^2}
 \biggl[
 \frac{3}{2}y_e^{}y_e^\dagger +Y_2
  -\frac{9}{4}g^2_1-\frac{9}{4}g_2^2
 \biggr]y_e^{} ,
 \end{align}
 where
 \begin{equation}
  Y_2\equiv {\rm Tr}(3y_u^{} y_u^\dagger +3y_d^{} y_d^\dagger 
 +y_e^{} y_e^\dagger )~.
 \end{equation}
Above the SUSY scale $Q_{\mathrm{SUSY}}$, the RGEs are 
\begin{align}
  \frac{d f_u^{}}{d \ln Q}&=\frac{1}{16\pi^2}
 \biggl[3{\rm Tr}(f_u^\dagger f_u^{})+3f_u^{} f_u^\dagger +f_d^{}f_d^\dagger
  -\frac{13}{15}g^2_1-3g_2^2-\frac{16}{3}
  g_3^2\biggr]f_u^{} ~, \nonumber \\
  \frac{d f_d^{}}{d \ln Q} &=\frac{1}{16\pi^2}
 \biggl[{\rm Tr}(3f_d^\dagger f_d^{}+f_e^\dagger f_e^{})
 +3f_d^{}f_d^\dagger +f_u^{}f_u^\dagger
  -\frac{7}{15}g^2_1-3g_2^2-\frac{16}{3}
  g_3^2\biggr]f_d^{} ~, \nonumber \\
  \frac{d f_e^{}}{d \ln Q} &=\frac{1}{16\pi^2}
 \biggl[{\rm Tr}(3f_d^\dagger f_d^{}+f_e^\dagger f_e^{})
 +3f_e f_e^\dagger  
  -\frac{9}{5}g^2_1-3g_2^2\biggr]f_e ~.
\end{align}
The relation between the Yukawa couplings above and below the SUSY scale, \textit{i.e.}, $f_k$ and $y_k$, are given in Sec.~\ref{sec:thcorr}.

%%%%%%%%%%%%%%%%%%%%%%%%%%%%%%%%%%
\subsection{Matching conditions}
\label{sec:thcorr}
%%%%%%%%%%%%%%%%%%%%%%%%%%%%%%%%%%%

There are two mass thresholds between the electroweak and GUT scales: the gaugino mass scale, $Q_{\mathrm{gaugino}}$, and the SUSY scale, $Q_{\mathrm{SUSY}}$. At the gaugino mass scale, we use the one-loop matching conditions for the gauge coupling constants: 
\begin{align}
  \frac{1}{g^2_1 (Q_{\mathrm{gaugino}}^+)} &= \frac{1}{g^2_1 (Q_{\mathrm{gaugino}}^-)} ~, \nonumber \\ 
  \frac{1}{g^2_2 (Q_{\mathrm{gaugino}}^+)} &= \frac{1}{g^2_2 (Q_{\mathrm{gaugino}}^-)} - \frac{1}{8\pi^2} \biggl(\frac{4}{3}\biggr) \ln \biggl(\frac{Q_{\mathrm{gaugino}}}{M_2}\biggr)  ~, \nonumber \\ 
  \frac{1}{g^2_3 (Q_{\mathrm{gaugino}}^+)} &= \frac{1}{g^2_3 (Q_{\mathrm{gaugino}}^-)} - \frac{1}{8\pi^2} \left( 2 \right) \ln \biggl(\frac{Q_{\mathrm{gaugino}}}{M_3}\biggr)  ~,
\end{align}
where the gauge couplings in the left(right)-hand side are those above (below) the gaugino mass scale. 

At the SUSY scale, we do not have threshold corrections for the gauge coupling constants since we have assumed that all of the SUSY particles except for gauginos have the same mass.\footnote{It is straightforward to relax this assumption. For instance, if higgsinos have a different mass from the other SUSY masses, the matching conditions for the gauge coupling constants are given by 
\begin{align}
  \frac{1}{g^2_1 (Q_{\mathrm{SUSY}}^+)} &= \frac{1}{g^2_1 (Q_{\mathrm{SUSY}}^-)} - \frac{1}{8\pi^2} \biggl(\frac{2}{5}\biggr) \ln \biggl(\frac{Q_{\mathrm{SUSY}}}{m_{\widetilde{H}}}\biggr) ~, \nonumber \\ 
  \frac{1}{g^2_2 (Q_{\mathrm{SUSY}}^+)} &= \frac{1}{g^2_2 (Q_{\mathrm{SUSY}}^-)} - \frac{1}{8\pi^2} \biggl(\frac{2}{3}\biggr) \ln \biggl(\frac{Q_{\mathrm{SUSY}}}{m_{\widetilde{H}}}\biggr)  ~, \nonumber \\ 
  \frac{1}{g^2_3 (Q_{\mathrm{SUSY}}^+)} &= \frac{1}{g^2_3 (Q_{\mathrm{SUSY}}^-)}   ~.
\end{align}
}
For the Yukawa couplings, we use the tree-level matching conditions: 
\begin{align}
  f_{u}(Q_{\mathrm{SUSY}}^+)&=\frac{1}{\sin\beta}{y}_{u}(Q_{\mathrm{SUSY}}^-)~, \nonumber \\
  f_{d}(Q_{\mathrm{SUSY}}^+)&=\frac{1}{\cos\beta}{y}_{d}(Q_{\mathrm{SUSY}}^-)~,  \nonumber \\
  f_{e}(Q_{\mathrm{SUSY}}^+)&=\frac{1}{\cos\beta}{y}_{e}(Q_{\mathrm{SUSY}}^-)~.
 \end{align}

%%%%%%%%%%%%%%%%%%%%%%%%%%%%%%%%%%%%%%%%%%%%%%
\section{Proton decay calculation}
\label{sec:protondecay}
\renewcommand{\theequation}{B.\arabic{equation}}
\setcounter{equation}{0}
%%%%%%%%%%%%%%%%%%%%%%%%%%%%%%%%%%%%%%%%%%%%%%

In this section we review the calculation of proton lifetimes in SU(5). We consider $p \to K^+ \bar{\nu}$ and $p \to e^+ \pi^0$ in Sec.~\ref{sec:dim5} and Sec.~\ref{sec:dim6}, respectively, which are the dominant decay channels in our setup. See Refs.~\cite{Goto:1998qg,Nagata:2013sba, Nagata:2013ive, Ellis:2015rya,Ellis:2019fwf} for more detailed discussions. We summarize the additional input parameters for proton decay calculation in Table~\ref{tab:proton}.

%%%%%%%%%%%%%%%%%%%%%%%%%%%%%%%%%%%%%%%%%%%%%%%%%%%%%%%%%%%%%%%%%%%%%%%%%%%%%%%5
\begin{table}[t]
  \centering
  \caption{Additional input parameters for proton decay calculation~\cite{ParticleDataGroup:2020ssz}. We use the Wolfenstein parameters (see Ref.~\cite{ParticleDataGroup:2020ssz} for their definition) to obtain the CKM matrix elements.  }
%  \vspace{3mm}
  \begin{tabular}{ll}
    \hline\hline
    Proton mass $m_p$ & $938.27208816(29)~\mathrm{MeV}$ \\ 
    Pion mass $m_{\pi}$ & $134.9768(5)~\mathrm{MeV}$ \\ 
    Kaon mass $m_{K}$ & $493.677(13)~\mathrm{MeV}$ \\
    Wolfenstein parameter $\lambda$ & $0.22500 \pm 0.00067$ \\
    Wolfenstein parameter $A$ & $0.826^{+0.018}_{-0.015}$ \\
    Wolfenstein parameter $\bar{\rho}$ & $0.159 \pm 0.010$ \\
    Wolfenstein parameter $\bar{\eta}$ & $0.348 \pm 0.010$ \\
    \hline\hline
  \end{tabular}
  \label{tab:proton}
\end{table}
%%%%%%%%%%%%%%%%%%%%%%%%%%%%%%%%%%%%%%%%%%%%%%%%%%%%%%%%%%%%%%%%%%%%%%%%%%%%%%%%

%%%%%%%%%%%%%%%%%%%%%%%%%%%%%%%%%%%
\subsection{$p \to K^+ \bar{\nu}$}
\label{sec:dim5}
%%%%%%%%%%%%%%%%%%%%%%%%%%%%%%%%%%%

The $p \to K^+ \bar{\nu}$ decay process is induced by the exchange of the color-triplet Higgs multiplet, whose effect can be described by the dimension-five effective operators: 
\begin{equation}
  {\cal L}_{5}^{\rm eff}
 =C_{5L}^{i j k l}{\cal O}^{5L}_{i j k l}
 +C_{5R}^{i j k l}{\cal O}^{5R}_{i j k l} + {\rm h.c.}
 ~,
 \label{eq:l5eff}
 \end{equation}
 with 
 \begin{align}
  {\cal O}^{5L}_{i j k l}&=\int d^2\theta ~
 \frac{1}{2}\epsilon_{abc}\epsilon_{mn}\epsilon_{pq}~
 Q^{am}_iQ^{bn}_j
 Q_k^{cp}L^q_l~,
  \nonumber\\
 {\cal O}^{5R}_{i j k l}&=\int d^2\theta ~
 \epsilon^{abc}
  \bigl(u^c_i\bigr)_a
  e^c _j
 \bigl(u^c _k\bigr)_b
 \bigl(d^c _l\bigr)_c ~,
\end{align}
where $i,j,\dots$ denote the generation indices, $a,b,c$ the color indices, and $p,q$ the $\mathrm{SU}(2)_L$ indices; $\epsilon_{abc}$ and $\epsilon_{pq}$ are the Levi-Civita symbols. The Wilson coefficients at the GUT scale are given by
\begin{align}
  C^{i j k l}_{5L}  (Q_{\rm G}) &= \frac{1}{M_{H_C}} f_{ui}e^{i\varphi_i}\delta^{ij}(V^*)^{kl}f_{dl} ~,
\nonumber \\
  C^{i j k l}_{5R} (Q_{\rm G}) & = \frac{1}{M_{H_C}} f_{ui} V^{ij}(V^*)^{kl}f_{dl}e^{-i\varphi_k} ~,
 \label{eq:dim5gutmatchijkl}
\end{align}
where $f_{ui}$ and $f_{dl}$ are the up- and down-type quark Yukawa couplings\footnote{We could have used the lepton Yukawa couplings $f_{el}$ instead of $f_{dl}$, since they are identical in SU(5). In practice, however, we obtain different values for $f_{dl}$ and $f_{el}$, running them from the electroweak scale up to the GUT scale. This difference may be accounted for by the effect of Planck-scale suppressed operators~\cite{Ellis:1979fg, Panagiotakopoulos:1984wf, Nath:1996qs, Nath:1996ft, Bajc:2002pg}. The uncertainty in the proton-decay calculation resulting from the choice of these Yukawa couplings is discussed in Ref.~\cite{Ellis:2019fwf}. It is found that our choice leads to a conservative estimate for the proton decay rate.  } at the GUT scale and $V^{ij}$ are the Cabibbo-Kobayashi-Maskawa (CKM) matrix elements. 
The phase factors $\varphi_i$, which are defined such that they satisfy the condition $\sum_i \varphi_i=0$, cannot be determined from low-energy observables. The uncertainty in the proton-lifetime calculation stemming from our ignorance of these phases is discussed in Sec.~\ref{sec:highscale}. It is found that only the operators $ {\cal O}^{5L}_{i i 1 j}$ and $ {\cal O}^{5R}_{331k}$ with $i=2,3$, $j=1,2,3$ and $k=1,2$ have sizable contributions.

From the GUT scale down to the SUSY scale, the RGEs for the Wilson coefficients are
\begin{align}
  \frac{d}{d\ln Q} C^{ijkl}_{5L} &=
 \frac{1}{16\pi^2}\biggl[-\frac{2}{5}g_1^2 -6g_2^2 -8g_3^2
 +f_{u_i}^2 +f_{d_i}^2 + f_{u_j}^2 +f_{d_j}^2 + f_{u_k}^2 + f_{d_k}^2 +
  f_{e_l}^2
 \biggr] C^{ijkl}_{5L}~, \nonumber \\
  \frac{d}{d\ln Q} C^{ijkl}_{5R} &=
 \frac{1}{16\pi^2}\biggl[-\frac{12}{5}g_1^2 -8g_3^2
 +2f_{u_i}^2 + 2f_{e_j}^2 + 2f_{u_k}^2 + 2f_{d_l}^2
 \biggr] C^{ijkl}_{5R}~.
\end{align}
At the SUSY breaking scale $Q_{\rm SUSY}$, sfermions are integrated out via the wino or higgsino exchange processes at one-loop, and these coefficients are matched onto the coefficients of the following effective operators 
\begin{align}
  {\cal L}^{\text{eff}}_{\rm SM}&=C^{\widetilde{H}}_i {\cal O}_{1i33}
 + C^{\widetilde{W}}_{jk}\widetilde{\cal O}_{1jjk}
 + C^{\widetilde{W}}_{jk}\widetilde{\cal O}_{j1jk}
 + \bar{C}^{\widetilde{W}}_{jk}\widetilde{\cal O}_{jj1k}
 ~,
\end{align}
with $i =1,2$, $j=2,3$, and $k=1,2,3$,
where the effective operators have the form
\begin{align}
 {\cal O}_{ijkl} &\equiv \epsilon_{abc}(u^a_{Ri}d^b_{Rj})
(Q_{Lk}^c \cdot L_{Ll}) ~, \nonumber \\
 \widetilde{\cal O}_{ijkl} &\equiv \epsilon_{abc} \epsilon^{\alpha\beta}
\epsilon^{\gamma\delta} (Q^a_{Li\alpha}Q^b_{Lj\gamma})
(Q_{Lk\delta}^c L_{Ll\beta}) ~.
\label{eq:effopPD}
\end{align}
We have 
\begin{align}
	C_i^{\widetilde{H}} (Q_{\rm SUSY}) &=\frac{f_tf_\tau}{(4\pi)^2} C^{*331i}_{5R}(Q_{\rm SUSY}) F(\mu, m_{\widetilde{t}_R}^2, m_{\widetilde{\tau}_R}^2) ~,\nonumber \\ 
	C_{jk}^{\widetilde{W}} (Q_{\rm SUSY}) &=\frac{\alpha_2}{4\pi} C^{jj1k}_{5L}(Q_{\rm SUSY}) \bigl[ F(M_2, m_{\widetilde{Q}_1}^2, m_{\widetilde{Q}_j}^2) +F(M_2, m_{\widetilde{Q}_j}^2, m_{\widetilde{L}_k}^2) \bigr] ~,\nonumber \\ 
	\bar{C}_{jk}^{\widetilde{W}} (Q_{\rm SUSY}) &= -\frac{3}{2}\frac{\alpha_2}{4\pi} C^{jj1k}_{5L}(Q_{\rm SUSY}) \bigl[ F(M_2, m_{\widetilde{Q}_j}^2, m_{\widetilde{Q}_j}^2) +F(M_2, m_{\widetilde{Q}_1}^2, m_{\widetilde{L}_k}^2) \bigr] ~,
\end{align}
where\footnote{Notice   
\begin{align}
  F(M, m^2, m^2) &= M \biggl[\frac{1}{m^2 - M^2} - \frac{M^2}{(m^2 - M^2)^2} \ln \biggl(\frac{m^2}{M^2}\biggr) \biggr]~, \qquad 
  F(m, m^2, m^2) = \frac{1}{2m} ~.
\end{align}
} 
\begin{equation}
	F(M, m_1^2, m_2^2) \equiv \frac{M}{m_1^2-m_2^2} \biggl[ \frac{m_1^2}{m_1^2-M^2}\ln \biggl( \frac{m_1^2}{M^2}\biggr) - \frac{m_2^2}{m_2^2-M^2}\ln \biggl( \frac{m_2^2}{M^2}\biggr) \biggr] ~. 
\end{equation}
 
From the SUSY breaking scale to the electroweak scale, the RGEs of Wilson coefficients are~\cite{Alonso:2014zka}
\begin{align}
	\frac{d}{d\ln Q} C_i^{\widetilde{H}} &= \frac{1}{16\pi^2} \biggl[ 
 		-\frac{11}{10}g_1^2 -\frac{9}{2}g_2^2 -4g_3^2 +\frac{1}{2}y_t^2 \biggr]   C_i^{\widetilde{H}} ~,\nonumber \\ 
	\frac{d}{d\ln Q} C_{jk}^{\widetilde{W}} &= \frac{1}{16\pi^2} \biggl[ 
 		-\frac{1}{5}g_1^2 -3g_2^2 -4g_3^2 +y_{u_j}^2 \biggr]   C_{jk}^{\widetilde{W}} -4\frac{g_2^2}{16\pi^2}\bigl[ 2C_{jk}^{\widetilde{W}} +\bar{C}_{jk}^{\widetilde{W}}  \bigr] ~,\nonumber \\ 
	\frac{d}{d\ln Q} \bar{C}_{jk}^{\widetilde{W}} &= \frac{1}{16\pi^2} \biggl[ 
 		-\frac{1}{5}g_1^2 -3g_2^2 -4g_3^2 +y_{u_j}^2 \biggr]   \bar{C}_{jk}^{\widetilde{W}} -4\frac{g_2^2}{16\pi^2}\bigl[ 2C_{jk}^{\widetilde{W}} +\bar{C}_{jk}^{\widetilde{W}}  \bigr] ~.
\end{align}
At the electroweak scale, these coefficients lead to the coefficients of the dimension-six operators
\begin{align}
  {\cal L}(p\to K^+\bar{\nu}_i^{})
 =&C_{RL}(usd\nu_i)\bigl[\epsilon_{abc}(u_R^as_R^b)(d_L^c\nu_i^{})\bigr]
 +C_{RL}(uds\nu_i)\bigl[\epsilon_{abc}(u_R^ad_R^b)(s_L^c\nu_i^{})\bigr]
 \nonumber \\
 +&C_{LL}(usd\nu_i)\bigl[\epsilon_{abc}(u_L^as_L^b)(d_L^c\nu_i^{})\bigr]
 +C_{LL}(uds\nu_i)\bigl[\epsilon_{abc}(u_L^ad_L^b)(s_L^c\nu_i^{})\bigr]
 ~,
 \end{align}
with 
\begin{align}
	C_{RL}(usd\nu_\tau ;m_t) &= -V_{td} C_2^{\widetilde{H}} (m_t)~,\nonumber \\ 
	C_{RL}(uds\nu_\tau ;m_t) &= -V_{ts} C_1^{\widetilde{H}} (m_t)~,\nonumber \\ 
	C_{LL}(usd\nu_k ;m_t) &= \sum_{j=2,3} V_{j1}V_{j2} C_{jk}^{\widetilde{W}} (m_t)~,\nonumber \\ 
	C_{LL}(uds\nu_k ;m_t) &= \sum_{j=2,3} V_{j1}V_{j2} C_{jk}^{\widetilde{W}} (m_t)~.
\end{align}

These coefficients are further run down to the hadronic scale $Q_{\rm had}=2 ~{\rm GeV}$, and the running effects are given by $A_L= C(Q_{\rm had})/C(m_t)$ as follows, where we use the two-loop QCD result given in Ref.~\cite{Nihei:1994tx}:
\begin{equation}
	A_L=  \begin{cases}
	\quad\biggl[
	\frac{\alpha_3(Q_{\rm had})}{\alpha_3(m_b)}
    	\biggr]^{\frac{6}{25}}
    	\biggl[
    	\frac{\alpha_3(m_b)}{\alpha_3(m_t)}
    	\biggr]^{\frac{6}{23}}
    	\biggl[
    	\frac{\alpha_3(Q_{\rm had})+\frac{50\pi}{77}}{\alpha_3(m_b)+\frac{50\pi}{77}}
    	\biggr]^{-\frac{2047}{11550}}
    	\biggl[
    	\frac{\alpha_3(m_b)+\frac{23\pi}{29}}{\alpha_3(m_t)+\frac{23\pi}{29}}
    	\biggr]^{-\frac{1375}{8004}} ~,\quad\quad {\rm for}~C_{LL} ~, \\
	\quad\biggl[
	\frac{\alpha_3(Q_{\rm had})}{\alpha_3(m_b)}
    	\biggr]^{\frac{6}{25}}
    	\biggl[
    	\frac{\alpha_3(m_b)}{\alpha_3(m_t)}
    	\biggr]^{\frac{6}{23}}
    	\biggl[
    	\frac{\alpha_3(Q_{\rm had})+\frac{50\pi}{77}}{\alpha_3(m_b)+\frac{50\pi}{77}}
    	\biggr]^{-\frac{173}{825}}
    	\biggl[
    	\frac{\alpha_3(m_b)+\frac{23\pi}{29}}{\alpha_3(m_t)+\frac{23\pi}{29}}
    	\biggr]^{-\frac{430}{2001}} ~,\quad\quad {\rm for}~C_{RL} ~.
	\end{cases}
  \label{sec:al}
\end{equation}
The partial decay widths of $p \to K^+ \bar{\nu}_i$ are now computed as 
\begin{equation}
 \Gamma (p\to  K^+ \bar{\nu}_i)=
\frac{m_p}{32\pi}\biggl(1-\frac{m_K^2}{m_p^2}\biggr)^2
\vert {\cal A}(p\to  K^+ \bar{\nu}_i) \vert^2 ~,
\end{equation}
where $m_p$ and $m_\pi$ denote the masses of the proton and pion, respectively, and 
\begin{align}
  \label{eq:ApKpnm}
       {\cal A}(p\to K^+\bar{\nu}_i)&=
  C_{RL}(usd\nu_i ;Q_{\rm had})\langle K^+\vert (us)_Rd_L\vert p\rangle
  +
  C_{RL}(uds\nu_i ;Q_{\rm had})\langle K^+\vert (ud)_Rs_L\vert p\rangle 
  \nonumber \\
  &+
  C_{LL}(usd\nu_i ;Q_{\rm had})\langle K^+\vert (us)_Ld_L\vert p\rangle
  +C_{LL}(uds\nu_i ;Q_{\rm had})\langle K^+\vert (ud)_Ls_L\vert p\rangle
  ~.
\end{align}
For the hadron matrix elements, we use the values obtained with QCD lattice simulations~\cite{Yoo:2021gql} 
\begin{align}
  {\langle K^+|(us)_Ld_L|p\rangle} &= 0.0284(30)(17)(12) ~{\rm GeV}^2 ~, \nonumber\\
  {\langle K^+|(ud)_Ls_L|p\rangle} &= 0.1006(80)(60)(46) ~{\rm GeV}^2 ~, \nonumber\\
  {\langle K^+|(us)_Rd_L|p\rangle} &= -0.0398(31)(20)(52) ~{\rm GeV}^2 ~, \nonumber\\
  {\langle K^+|(ud)_Rs_L|p\rangle} &= -0.109(10)(8)(14) ~{\rm GeV}^2 ~,
\end{align}
where the first, second, and third parentheses show the statistical error, the systematic error caused by excited states, and the uncertainty due to the continuum extrapolation.

%%%%%%%%%%%%%%%%%%%%%%%%%%%%%%%%%%%
\subsection{$p \to e^+ \pi^0$}
\label{sec:dim6}
%%%%%%%%%%%%%%%%%%%%%%%%%%%%%%%%%%%

The $p \to e^+ \pi^0$ decay mode is induced by the SU(5) gauge boson exchange, whose contribution can be described by the dimension-six operators 
\begin{equation}
  {\cal L}_{6}^{\rm eff}
 =C_{6(1)}^{i j k l}{\cal O}^{6(1)}_{i j k l}
 +C_{6(2)}^{i j k l}{\cal O}^{6(2)}_{i j k l}
 ~,
 \label{eq:l6eff}
 \end{equation}
 where 
 \begin{align}
  {\cal O}^{6(1)}_{i j k l}&=\int d^2\theta d^2\bar{\theta}~
 \epsilon_{abc}\epsilon_{mn}
 \bigl(u^{c\dagger}_i\bigr)^a
 \bigl(d^{c\dagger} _j\bigr)^b
 e^{-\frac{2}{3}g^\prime B}
 \bigl(e^{2g_3G}Q_k^m\bigr)^cL^n_l~,
  \nonumber\\
 {\cal O}^{6(2)}_{i j k l}&=\int d^2\theta d^2\bar{\theta}~
 \epsilon_{abc}\epsilon_{mn}~
 Q^{am}_iQ^{bn}_j
 e^{\frac{2}{3}g^\prime B}
 \bigl(e^{-2g_3G}u^{c\dagger} _k\bigr)^c
 e^{c\dagger} _l~,
 \end{align}
 with $G$ and $B$ the SU(3)$_C$ and U(1)$_Y$ gauge vector superfields,
 respectively.
 The Wilson coefficients are given by
\begin{align}
  C^{i j k l}_{6(1)}  (Q_{\rm G}) &= -\frac{g_5^2}{M_X^2} e^{i\varphi_i}\delta^{ik}\delta^{jl}  ~,
\nonumber \\
  C^{i j k l}_{6(2)} (Q_{\rm G}) & = -\frac{g_5^2}{M_X^2} e^{i\varphi_i}\delta^{ik} (V^*)^{jl} ~.
 \label{eq:dim6gutmatch}
\end{align}
 
The one-loop RGEs\footnote{The two-loop RGEs are also available in Ref.~\cite{Hisano:2013ege}.} from the GUT scale down to the SUSY breaking scale are~\cite{Munoz:1986kq}
\begin{align}
 \frac{d}{d\ln Q}C^{ijkl}_{6(1)} &=\biggl[ 
 	\frac{\alpha_1}{4\pi}\biggl( -\frac{11}{15}\biggr) +\frac{\alpha_2}{4\pi}\left( -3\right) +\frac{\alpha_3}{4\pi}\biggl( -\frac{8}{3}\biggr)
 \biggr] C^{ijkl}_{6(1)} ~, \nonumber \\
 \frac{d}{d\ln Q}C^{ijkl}_{6(2)} &=\biggl[ 
 	\frac{\alpha_1}{4\pi}\biggl( -\frac{23}{15}\biggr) +\frac{\alpha_2}{4\pi}\left( -3\right) +\frac{\alpha_3}{4\pi}\biggl( -\frac{8}{3}\biggr)
 \biggr] C^{ijkl}_{6(2)} ~,
\end{align}
which give
 \begin{align}
 C^{i j k l}_{6(1)}  (Q_{\rm SUSY})=  \biggl[
    \frac{\alpha_3(Q_{\rm SUSY})}{\alpha_3(Q_{\rm G})}
    \biggr]^{\frac{4}{9}}
    \biggl[
    \frac{\alpha_2(Q_{\rm SUSY})}{\alpha_2(Q_{\rm G})}
    \biggr]^{-\frac{3}{2}}
    \biggl[
    \frac{\alpha_1(Q_{\rm SUSY})}{\alpha_1(Q_{\rm G})}
    \biggr]^{-\frac{1}{18}} 
    C^{i j k l}_{6(1)} (Q_{\rm G})  ~, \nonumber \\
  C^{i j k l}_{6(2)}  (Q_{\rm SUSY})=  \biggl[
    \frac{\alpha_3(Q_{\rm SUSY})}{\alpha_3(Q_{\rm G})}
    \biggr]^{\frac{4}{9}}
    \biggl[
    \frac{\alpha_2(Q_{\rm SUSY})}{\alpha_2(Q_{\rm G})}
    \biggr]^{-\frac{3}{2}}
    \biggl[
    \frac{\alpha_1(Q_{\rm SUSY})}{\alpha_1(Q_{\rm G})}
    \biggr]^{-\frac{23}{198}} 
    C^{i j k l}_{6(2)} (Q_{\rm G})  ~.
 \end{align}
 
Below the SUSY breaking scale, the RGEs of Wilson coefficients are given by~\cite{Abbott:1980zj}
 \begin{align}
 \frac{d}{d\ln Q}C^{ijkl}_{6(1)} &=\biggl[ 
 	\frac{\alpha_1}{4\pi}\biggl( -\frac{11}{10}\biggr) +\frac{\alpha_2}{4\pi}\biggl( -\frac{9}{2}\biggr) +\frac{\alpha_3}{4\pi}\left( -4\right)
 \biggr] C^{ijkl}_{6(1)} ~, \nonumber \\
 \frac{d}{d\ln Q}C^{ijkl}_{6(2)} &=\biggl[ 
 	\frac{\alpha_1}{4\pi}\biggl( -\frac{23}{10}\biggr) +\frac{\alpha_2}{4\pi}\biggl( -\frac{9}{2}\biggr) +\frac{\alpha_3}{4\pi}\left( -4\right)
 \biggr] C^{ijkl}_{6(2)} ~.
\end{align}
From the SUSY scale to the gaugino-mass scale, these RGEs give the renormalization factors as 
  \begin{align}
 C^{i j k l}_{6(1)}  (Q_{\rm gaugino})=  \biggl[
    \frac{\alpha_3(Q_{\rm gaugino})}{\alpha_3(Q_{\rm SUSY})}
    \biggr]^{\frac{2}{5}}
    \biggl[
    \frac{\alpha_2(Q_{\rm gaugino})}{\alpha_2(Q_{\rm SUSY})}
    \biggr]^{\frac{27}{22}}
    \biggl[
    \frac{\alpha_1(Q_{\rm gaugino})}{\alpha_1(Q_{\rm SUSY})}
    \biggr]^{-\frac{11}{82}} 
    C^{i j k l}_{6(1)} (Q_{\rm SUSY})  ~, \nonumber \\
  C^{i j k l}_{6(2)}  (Q_{\rm gaugino})=  \biggl[
    \frac{\alpha_3(Q_{\rm gaugino})}{\alpha_3(Q_{\rm SUSY})}
    \biggr]^{\frac{2}{5}}
    \biggl[
    \frac{\alpha_2(Q_{\rm gaugino})}{\alpha_2(Q_{\rm SUSY})}
    \biggr]^{\frac{27}{22}}
    \biggl[
    \frac{\alpha_1(Q_{\rm gaugino})}{\alpha_1(Q_{\rm SUSY})}
    \biggr]^{-\frac{23}{82}} 
    C^{i j k l}_{6(2)} (Q_{\rm SUSY})  ~,
 \end{align}
while from the gaugino-mass scale to the electroweak scale, we have 
  \begin{align}
 C^{i j k l}_{6(1)}  (m_t)=  \biggl[
    \frac{\alpha_3(m_t)}{\alpha_3(Q_{\rm gaugino})}
    \biggr]^{\frac{2}{7}}
    \biggl[
    \frac{\alpha_2(m_t)}{\alpha_2(Q_{\rm gaugino})}
    \biggr]^{\frac{27}{38}}
    \biggl[
    \frac{\alpha_1(m_t)}{\alpha_1(Q_{\rm gaugino})}
    \biggr]^{-\frac{11}{82}} 
    C^{i j k l}_{6(1)} (Q_{\rm gaugino})  ~, \nonumber \\
  C^{i j k l}_{6(2)}  (m_t)=  \biggl[
    \frac{\alpha_3(m_t)}{\alpha_3(Q_{\rm gaugino})}
    \biggr]^{\frac{2}{7}}
    \biggl[
    \frac{\alpha_2(m_t)}{\alpha_2(Q_{\rm gaugino})}
    \biggr]^{\frac{27}{38}}
    \biggl[
    \frac{\alpha_1(m_t)}{\alpha_1(Q_{\rm gaugino})}
    \biggr]^{-\frac{23}{82}} 
    C^{i j k l}_{6(2)} (Q_{\rm gaugino})  ~.
 \end{align} 
 
At the electroweak scale, these Wilson coefficients are matched onto the coefficients of the operators 
\begin{align}
  \mathcal{L}(p \to \pi^0 \ell_i^+) &= C_{RL} (udu\ell_i) [\epsilon_{abc} (u_R^a d_R^b) (u_L^c \ell_{Li})] + C_{LR} (udu\ell_i)
  [\epsilon_{abc} (u_L^a d_L^b) (u_R^c \ell_{Ri})] ~,
\end{align}
as 
\begin{align}
  C_{RL}(udu\ell_i ;m_t)&=C^{111i}_{6(1)}(m_t)~, \nonumber \\
  C_{LR}(udu\ell_i ;m_t)&=V_{j1}\bigl[
 C^{1j1i}_{6(2)}(m_t)+C^{j11i}_{6(2)}(m_t)
 \bigr]~.
\end{align}
These coefficients are run down to the hadronic scale according to the QCD RGEs. The renormalization factor is given in the second row in Eq.~\eqref{sec:al}. By using the Wilson coefficients at the scale $Q_{\mathrm{had} } = 2~\mathrm{GeV}$, we compute the partial decay width of $p \to \pi^0 e^+$ as 
\begin{equation}
 \Gamma (p\to  \pi^0 e^+)=
\frac{m_p}{32\pi}\biggl(1-\frac{m_\pi^2}{m_p^2}\biggr)^2
\bigl[
\vert {\cal A}_L \vert^2+
\vert {\cal A}_R \vert^2
\bigr]~,
\end{equation}
where
\begin{align}
 {\cal A}_L&=
C_{RL}(udu\ell_1 ;Q_{\rm had})\langle \pi^0\vert (ud)_Ru_L\vert p\rangle
~,\nonumber \\
 {\cal A}_R&=
C_{LR}(udu\ell_1 ;Q_{\rm had})\langle \pi^0\vert (ud)_Ru_L\vert p\rangle
~,
\end{align}
and the matrix element is given by~\cite{Yoo:2021gql} 
\begin{align}
  {\langle \pi^0|(ud)_Ru_L|p\rangle} &= -0.112(11)(14)(18) ~{\rm GeV}^2 ~.
\end{align}

%%%%%%%%%%%%%%%%%%%%%%%%%%%%%%%%%%%%%%%

\bibliographystyle{utphysmod}
\bibliography{ref}

%%%%%%%%%%%%%%%%%%%%%%%%%%%%%%%%%%%%%%%

\end{document}